\documentclass[12pt,a4paper]{article}
\usepackage{graphicx}
\usepackage{amsmath}
\usepackage{flafter}

\input{tcilatex}

\begin{document}

\title{Pole dynamics for the Flierl-Petviashvili equation and zonal flow }
\author{F. Spineanu$^{1}$, M. Vlad$^{1}$, K. Itoh$^{1}$, H. Sanuki$^{1}$ and S.-I.
Itoh$^{2}$ \\
$^{1}$ National Institute for Fusion Science\\
322-6 Oroshi-cho, Toki-shi, Gifu-ken 509-5292, Japan\\
$^{2}$ Research Institute for Applied Mechanics, Kyushu University, \\
Kasuga 816-8580, Japan}
\date{}
\maketitle

\begin{abstract}
We use a systematic method which allows us to identify a class of exact
solutions of the Flierl-Petvishvili equation. The solutions are periodic and
have one dimensional geometry. We examine the physical properties and find
that these structures can have a significant effect on the zonal flow
generation.

\end{abstract}

\section{Introduction}

The problem of coherent structures has been extensively investigated in
connection with the drift wave modes in tokamak. Numerical simulations and
experimental observations have provided strong evidence of the presence of
long-lived, cuasi-coherent structures even in deep turbulent regimes. The
theoretical models (close to similar descriptions in the physics of fluids,
atmosphere and ocean) emphasize the role of two types of nonlinearity :
scalar (or Korteweg deVries-type) and vectorial (convection of the
vorticity). Both are able to support vortical flow in a form of coherent and
long lived structures. The problems of generation from initial conditions
and the stability of monopolar or multipolar vortices are subjects of
intense research and one of the notable result is that not all of these
structures can be expected to be solitons in the sense of the Inverse
Scattering Method. Usually they are called solitary waves or solitary
vortices. A review, mainly oriented to plasma physics applications, has been
done by Horton and Hasegawa \cite{HH}.

The ion drift wave equation has a distinct dynamical character according to
the space-time scales involved. For shorter scales the Hasegawa-Mima-Charney
equation is obtained and the dynamics exhibits dipolar structures on the
Larmor radius scale. On larger scales the scalar (or KdV-type) nonlinearity
is prevailing and the structures are monopolar. Both are not solitonic but
very robust and long lived. In the latter case the one-dimensional version
of the equation can be reduced to the modified Korteweg De Vries (mKdV)
equation \cite{Ablowitz} and has been derived in various contexts in plasma
physics. Tasso \cite{Tasso}\ and later Petviashvili \cite{Petv1} have
derived versions of this equations applicable to the case where there is a
strong temperature gradient. It has been shown \cite{Lakhin}, \cite
{Spatschek1} that the equation proposed was in fact exclusively dependent of
the density gradient. Later the equation has been rederived along with a
careful analysis of the scales involved \cite{Spatschek2}, resolving a
controversy on the role of the temperature gradient. In the study of ocean
flows Flierl \cite{Flierl} has independently formulated an equation with the
same structure. A one dimensional version of the equation has been solved by
Lakhin et al. on an infinite domain \cite{Lakhin}, obtaining as solution the
KdV soliton. The two dimensional case has been examined by Kadomtsev and
Petviashvili \cite{KP} and by Petviashvili \emph{et al.} \cite{Petv2} using
trial functions to extremize a functional derived from the equation. The
solution they found is vortical monopolar. Boyd and Tan \cite{Boyd1} have
found a monopolar solution expressed as a sum of $45$ terms depending on the
radial coordinate $r$ through trigonometric functions. They have also proved
the non-existence of stable non-axisymmetric solutions to the stationary
Flierl-Petviashvili equation \cite{Boyd2}. There is a strong connection
between the stationary Flierl-Petviashvili equation and the
Zakharov-Kuznetsov equation \cite{ZK}, a two-dimensional generalization of
the KdV equation. This has also been investigated in Refs.\cite{TS} and \cite
{SS} where multidimensional nonlinear wave structures (of the electrostatic
drift wave branch) in inhomogeneous plasmas were studied by combining the
eigenvalue problem in inhomogeneity and the $2D$ nonlinear vortex equation.
For a two-dimensional generalization of the KdV equation numerical results
are available \cite{Iwasaki}. For shorter reference it is used also the name
Regularized Long Wave equation, RLW, since the equation proposed by
Peregrine \cite{Peregrine} can be brought to the same form.

The stationary two-dimensional equation derived in the studies mentioned
above has a very simple structure. and a one-dimensional solution is known,
the KdV soliton. We will examine the possible extension to an analytical
closed form of a two dimensional solution. We are motivated by the studies
of plasma ion mode instabilities, in particular the generation of radially
localised layers of sheared flow (zonal flow).

\section{Derivation of the equation}

The ion instabilities are dominated by the nonlinearity related to the
polarization drift. When the temperature gradient is very small the
quasi-three dimensional geometry assumed in the derivation of the
Haswgawa-Mima equation leads to suppression of the classical convection
nonlinearity although the presence of the later can still be important in a
fluctuation model. The ion polarisation drift nonlinearity may support
vortical flows of which certain states are stable and coherent (actually not
solitonic). The derivation of a nonlinear equation in this case starts by
the ion density continuity 
\begin{equation}
\left( \frac{\partial }{\partial t}+\mathbf{v}_{i\perp }\cdot \mathbf{\nabla 
}_{\perp }\right) n_{i}+n_{0}\left( \mathbf{\nabla }_{\perp }\cdot \mathbf{v}%
\right) =0  \label{icon}
\end{equation}
and assuming neutrality and adiabatic response of the particles 
\begin{equation*}
n_{i}\approx n_{e}\approx n_{0}\frac{e\varphi }{T_{e}}
\end{equation*}
The ion velocity in two dimension is 
\begin{equation}
\mathbf{v}_{i\perp }=\frac{-\mathbf{\nabla }_{\perp }\varphi \times \widehat{%
\mathbf{n}}}{B}-\frac{1}{\Omega _{i}B}\left( \frac{\partial }{\partial t}+%
\frac{-\mathbf{\nabla }_{\perp }\varphi \times \widehat{\mathbf{n}}}{B}\cdot 
\mathbf{\nabla }_{\perp }\right) \mathbf{\nabla }_{\perp }\varphi
\label{ivit}
\end{equation}
The Eqs.(\ref{icon}) and (\ref{ivit}) lead to a nonlinear equation for the
electrostatic potential. However, in these two equations there is still much
freedom and a detailed analysis of the time and space scales leads to
different particular forms of this equation. This is a multiple space and
time scale analysis \cite{MKT}, \cite{Spatschek2}, \cite{Su} 
\begin{eqnarray*}
x_{i} &=&\varepsilon ^{i}x \\
t_{i} &=&\varepsilon ^{i}t
\end{eqnarray*}
for $\varepsilon \ll 1$. It is possible to change to a moving frame, by
introducing a translation velocity, $u$. 
\begin{equation*}
\eta =y-ut
\end{equation*}
The speed $u$ can also be a function on certain space and time scales.

We recall the main steps of the analysis performed in Ref. \cite{Spatschek2}%
. In the scaling 
\begin{equation*}
\varphi =\varepsilon \varphi _{1}+\varepsilon ^{2}\varphi _{2}+\ldots
\end{equation*}
with 
\begin{equation*}
\varphi _{1}=\varphi _{1}\left( x_{0},x_{1},x_{2},...,\eta _{0},\eta
_{1},\eta _{2},...,t_{0},t_{1},t_{2},...\right)
\end{equation*}
\begin{equation*}
u\sim u_{1}\equiv O\left( \varepsilon \right)
\end{equation*}
\begin{equation*}
n_{0}=n_{0}\left( x_{1},x_{2},...\right)
\end{equation*}
\begin{equation*}
T_{e}=T_{e}\left( x_{1},x_{2},...\right)
\end{equation*}
In this scaling $n_{0}$ and $T_{e}$ have variations on only ``large'' space
scales, and the potential (which is of $\varepsilon $ -small amplitude)
varies on the small space scale, $x_{0}$ and slow time scale, $t_{0}$ and,
of course, on higher scales. The characteristic space scale for the
potential is then $\rho _{s}$ which is the spatial extension of the dipolar
vortex. The normalisation of the parameters can be done according to these
scales 
\begin{eqnarray*}
\phi &=&\frac{e\varphi }{T_{0}}\;,\;T=\frac{T_{e}}{T_{0}}\;,\;v=\frac{%
v_{i\perp }}{c_{s}} \\
x^{\prime } &=&x/\rho _{s}\;,\;\eta ^{\prime }=\eta /\rho _{s}\;,\;t^{\prime
}=\Omega _{i}^{-1}t
\end{eqnarray*}
and dropping the primes, one obtains 
\begin{eqnarray*}
&&\frac{\partial }{\partial t_{1}}\left[ \frac{1}{T\left( x_{1}\right) }-%
\mathbf{\nabla }_{\perp 0}^{2}\right] \phi _{1} \\
&&+u_{1}\frac{\partial }{\partial \eta _{1}}\mathbf{\nabla }_{\perp
0}^{2}\phi _{1}-u_{1}\left( \frac{1}{T\left( x_{1}\right) }+\frac{\kappa
_{n}\left( x_{1}\right) }{u_{1}}\right) \frac{\partial }{\partial \eta _{0}}%
\phi _{1} \\
&&-\left[ \left( -\mathbf{\nabla }_{\perp 0}\phi _{1}\times \widehat{\mathbf{%
n}}\right) \cdot \mathbf{\nabla }_{\perp 0}\right] \mathbf{\nabla }_{\perp
0}^{2}\phi _{1} \\
&=&0
\end{eqnarray*}
where 
\begin{equation*}
\kappa _{n}\equiv L_{n}^{-1}=\frac{1}{n_{0}}\frac{\partial n_{0}}{\partial x}
\end{equation*}
is varying on the larger space scale, $x_{1}$.

This is the Hasegawa-Mima equation governing the space variation of the
potential $\phi _{1}$ on the smaller space scale, $x_{0}$ and $\eta _{0}$ of
the order $\rho _{s}$ and on times of the scale $t_{1}$. On these scales
several simplifications are obvious, since the temperature and the density
gradient lengths can be taken constants 
\begin{eqnarray*}
\kappa _{n}/u_{1} &\sim &\text{constant on }x_{0} \\
T &\sim &\text{constant on }x_{0}
\end{eqnarray*}

A different dynamical equation is obtained on other space-time scale 
\begin{equation*}
\phi =\varepsilon ^{2}\phi _{2}+\varepsilon ^{3}\phi _{3}+...
\end{equation*}
\begin{equation*}
\phi _{2}\equiv \phi _{2}\left( x_{1},x_{2},...,\eta _{1},\eta
_{2},...,t_{5},t_{6},...\right)
\end{equation*}
\begin{equation*}
n_{0}=n_{0}\left( x_{2},x_{3},...\right)
\end{equation*}
\begin{equation*}
u=u_{2}\sim O\left( \varepsilon ^{2}\right)
\end{equation*}
\begin{equation*}
T=T\left( x_{2},x_{3},...\right)
\end{equation*}
The following combination has to be of a certain order in $\varepsilon $%
\begin{equation*}
\frac{1}{T\left( x_{2}\right) }+\frac{\kappa _{n}\left( x_{2}\right) }{u_{2}}%
\sim O\left( \varepsilon ^{2}\right)
\end{equation*}
Then the equation on the time scale of order $5$ is 
\begin{eqnarray*}
&&\frac{\partial }{\partial t_{5}}\frac{\phi _{2}}{T\left( x_{2}\right) }%
+u_{2}\frac{\partial }{\partial \eta _{1}}\mathbf{\nabla }_{\perp 1}^{2}\phi
_{2} \\
&&-u_{2}\left[ \frac{1}{T\left( x_{2}\right) }+\frac{\kappa _{n}\left(
x_{2}\right) }{u_{2}}\right] \frac{\partial \phi _{2}}{\partial \eta _{1}} \\
&&+\kappa _{T}\left( x_{2}\right) \phi _{2}\frac{\partial \phi _{2}}{%
\partial \eta _{1}} \\
&=&0
\end{eqnarray*}
The operator of Laplacean in two dimension acts on the larger space scales 
\begin{equation*}
\mathbf{\nabla }_{\perp 1}^{2}\equiv \frac{\partial ^{2}}{\partial x_{1}^{2}}%
+\frac{\partial ^{2}}{\partial \eta _{1}^{2}}
\end{equation*}
The dominant space variation is here on the scale 
\begin{equation*}
x_{1}=\frac{\rho _{s}}{\varepsilon }
\end{equation*}
which is much larger than the dipolar vortex scale. The condition imposed by
this ordering is 
\begin{equation*}
\kappa _{T}=\frac{\partial \kappa _{n}}{\partial x_{1}}\frac{1}{u_{2}}
\end{equation*}

In this range the nonlinear equation is dominated by the scalar
nonlinearity. This equation leads, after one integration over the $\eta _{1}$
coordinate, to the Flierl-Petviashvili equation.

Spatschek \emph{et al}. \cite{Spatschek2} also discuss intermediate
scalings, where both the scalar and the vectorial nonlinearities are
present. The stability analysis implies the concept of structural stability,
where the dynamics governed by one of the equations obtained above
(Hasegawa-Mima or scalar nonlinear) is perturbed with a term that is of the
other type. It has been shown that the dipolar vortices are broken into
separate monopolar vortices, while the monopolar vortices are structurally
stable. To these consideration one should also add the numerical results
from the collision of monopolar vortices of the Flierl-Petviashvili equation
(or Zakharov-Kusnetsov) \cite{Boyd1}, \cite{Boyd2}, \cite{Iwasaki}. The
monopolar vortices are stable and are not destroyed by collisions, however
the form and the amplitides are perturbed, showing again that they are not
exact solitons.

\section{The scalar nonlinearity equation}

In plasma physics applications, the Flierl-Petviashvili equation has the
form 
\begin{equation}
\Delta \phi =\alpha \phi -\beta \phi ^{2}  \label{eqorig}
\end{equation}
where $\alpha $ and $\beta $ are physical parameters, \emph{i.e.} functions
depending on $\left( x,y\right) $. This equation is obtained as the
stationary version of the equation which has been expressed in the system of
reference moving with the velocity $u$%
\begin{equation*}
y\rightarrow y-ut
\end{equation*}
In the following, the coordinates $x$ and $y$ are the coordinates in the
moving system. It is assumed that $y$ is the poloidal direction and $x$ is
the radial direction in tokamak.

In the tokamak context the equation has been frequently simplified 
\begin{equation*}
\frac{\partial ^{2}}{\partial x^{2}}\ll \frac{\partial ^{2}}{\partial y^{2}}
\end{equation*}
leading to 
\begin{equation}
\frac{\partial ^{2}\phi }{\partial y^{2}}=\alpha \phi -\beta \phi ^{2}
\label{onedim}
\end{equation}
whose solution (Lakhin \emph{et al}.\cite{Lakhin}) is 
\begin{equation}
\phi \left( x,y\right) =\frac{3\alpha /(2\beta )}{\cosh ^{2}\left( \frac{%
\sqrt{\alpha }}{2}y\right) }  \label{sollakh}
\end{equation}
The dependence on $x$ is only parametric here, via the coefficients $\alpha $
and $\beta $. This has the same form as the KdV soliton solution.

Assuming that $\alpha $ and $\beta $ are constants, this solution can be
extended indefinitely in the transversal ($x$) direction, as a ridge
propagating in the $y$ direction whose section is given by Eq.(\ref{sollakh}%
). If the plasma is homogeneous, this ridge can be rotated arbitrarly in
plane. We have then a family of quasi two dimensional solutions to the Eq.(%
\ref{eqorig}).

We will discuss the possibility to find other solutions of the \emph{%
two-dimensional} version of the equation.

\section{Starting from the one dimensional problem}

We can directly integrate the one dimensional version of the equation (\ref
{eqorig}). Actually, the solution (\ref{sollakh}) is one of the possible
solutions arising from the direct integration. For a reason that will become
clear later, we write the one dimensional version of the equation (\ref
{eqorig}) in the form 
\begin{equation*}
\frac{d^{2}\phi }{d\left( i\eta \right) ^{2}}=\alpha \phi -\beta \phi ^{2}
\end{equation*}
where $i\eta $ can be $y$ as in Eq.(\ref{onedim}). Multiplying by $d\phi
/d\eta $ and integrating once we have 
\begin{equation*}
-\frac{1}{2}\left( \frac{d\phi }{d\eta }\right) ^{2}=\frac{\alpha }{2}\phi
^{2}-\frac{\beta }{3}\phi ^{3}+\kappa
\end{equation*}
where $\kappa $ is a constant. Then we have the integral 
\begin{equation*}
\eta =\int^{\phi }\frac{d\phi }{\sqrt{\frac{2\beta }{3}\phi ^{3}-\alpha \phi
^{2}-2\kappa }}
\end{equation*}
The integral is elliptic and its inverse $\phi \left( \eta \right) $ has a
closed analytical expression for all $\kappa $. Particular expressions can
be written according to the reality of the roots of the third degree
polynomial under the square root. For $\kappa $ such that all roots ($a>b>c$%
) are real, we have, for $\phi >a$%
\begin{eqnarray*}
\int_{\phi }^{\infty }\frac{d\phi }{\sqrt{\left( \phi -a\right) \left( \phi
-b\right) \left( \phi -c\right) }} &=&g\mathrm{sn}^{-1}\left( \sin \varphi
,k\right) \\
&=&gF\left( \varphi ,k\right)
\end{eqnarray*}
where $\mathrm{sn}$ is the Jacobi elliptic sinus, $F$ is the incomplete
elliptic integral of the first kind and the notations are 
\begin{eqnarray*}
\varphi &=&\arcsin \sqrt{\frac{a-c}{\phi -c}} \\
g &=&\frac{2}{\sqrt{a-c}} \\
k^{2} &=&\frac{b-c}{a-c}
\end{eqnarray*}
This expression can be easily inverted to obtain $\phi $ as function of $%
\eta $. An analoguous result is obtained when $\kappa $ is such that the
root $a$ is real and $b$ and $c$ are complex. The integral is written, for $%
\phi >a$%
\begin{eqnarray*}
\int_{\phi }^{\infty }\frac{d\phi }{\sqrt{\left( \phi -a\right) \left[
\left( \phi -b_{1}\right) ^{2}+a_{1}^{2}\right] }} &=&g\mathrm{cn}%
^{-1}\left( \cos \varphi ,k\right) \\
&=&gF\left( \varphi ,k\right)
\end{eqnarray*}
where 
\begin{eqnarray*}
b_{1} &=&\frac{\left( b+\overline{b}\right) }{2},\;a_{1}^{2}=-\frac{\left( b-%
\overline{b}\right) ^{2}}{4} \\
g &=&\frac{1}{\sqrt{A}},\;A^{2}=\left( b_{1}-a\right) ^{2}+a_{1}^{2} \\
k^{2} &=&\frac{A+b_{1}-a}{2A} \\
\varphi &=&\arccos \frac{\phi -a-A}{\phi -a+A}
\end{eqnarray*}
Deatils on these Jacobian elliptic functions can be found in Ref.\cite{Byrd}.

\section{Constructing the solution from singularities}

In view of the application to the description of plasma drift wave, a
two-dimensional solution would be very useful. We already dispose of the
one-parameter family of solutions with geometry of fronts, which has been
obtained by simply translating a one-dimensional solution along the
transversal direction. We examine a class of solutions that can be seen\ as
an extension from the one-dimensional ones. We propose a more systematic
approach of construction that makes more explicit the nature of the
restriction leading to this quasi-one dimensional geometry.

We will start from the one dimensional model, taking the coefficients
constants. The only analytical formula for a solution of (\ref{eqorig}) is 
\cite{Lakhin} 
\begin{equation*}
\phi \left( y\right) =\phi _{0}\mathrm{sech}^{2}\left( \gamma y\right)
\end{equation*}
where 
\begin{eqnarray*}
\phi _{0} &=&\frac{3\alpha }{2\beta } \\
\gamma &=&\frac{\sqrt{\alpha }}{2}
\end{eqnarray*}
(note that in Lakhin \emph{et al}. the coefficient must be corrected by
dividing with $2$). We will use the following relation 
\begin{equation*}
\mathrm{sech}^{2}\left( z\right) =-\mathrm{cosech}^{2}\left( z-\frac{i\pi }{2%
}\right)
\end{equation*}
and the expression of the $\mathrm{cosech}$ function as a series using its
pole singularities. Then we obtain 
\begin{equation}
\phi \left( y\right) =\phi _{0}\mathrm{sech}^{2}\left( \gamma y\right)
=-\phi _{0}\sum_{l=-\infty }^{\infty }\frac{1}{\left( \gamma y-\frac{i\pi }{2%
}+il\pi \right) ^{2}}  \label{sech2}
\end{equation}

We are looking for solutions $\phi \left( x,y\right) $ of the $2D$ equation
and we try to build them from the motion of the pole singularities of the
one-dimensional solution. More specifically, we will make an ansatz for the
form of the $2D$ solution based on a particular choice of poles in the
complex $y$ plane, as suggested by Eq.(\ref{sech2}). We will assume that the
poles have positions that depend on the other coordinate, $x$. Imposing that
the function constructed in this way verifies the equation we obtain a set
of differential equations and constraint conditions for the positions of the
poles in the complex $y$ plane. The poles evolve with $x$ as time-like
variable. Expressed in other terms, we look for a function that is
meromorphic in the complex $y$ plane and whose poles depend on $x$. We find,
following other similar approaches, that the solution is an elliptic
function, \emph{i.e.} a doubly periodic meromorphic function of $y$, for all
values of $x$.

This procedure has been deveoped in the context of the exactly integrable
differential equations like Kortweg de Vries. We follow closely the methods
exposed in papers of Choodnovsky \cite{Chood}, Thickstun \cite{Thickstun}.
The work of Deconinck and Segur \cite{Decon} is particularly relevant for
our problem.

The following ansatz is suggested by the theory of $\tau $-functions in the
integrable equations context \cite{Mulase} 
\begin{equation}
\phi \left( x,y\right) =2\frac{\partial ^{2}}{\partial y^{2}}\ln \tau \left(
x,y\right)  \label{phitau}
\end{equation}
where 
\begin{equation}
\tau \left( x,y\right) =c\prod_{k=1}^{\infty }\left( 1-\frac{y}{y_{k}}%
\right) \exp \left( \frac{y}{y_{k}}\right)  \label{tau}
\end{equation}
On the other hand, there is also the suggestion from Eq.(\ref{sech2}) that
the function $\phi $ is periodic on the imaginary $y$ axis. We choose a
periodicity $iD$ (compare with $i\pi $ in the above equation) and take $N$
poles in each domain of periodicity. Then $\tau $ is 
\begin{equation}
\tau \left( x,y\right) \sim c\prod_{n=1}^{N}\prod_{l=-\infty }^{\infty }%
\frac{1}{\left( y-y_{n}-ilD\right) ^{2}}  \label{tausing}
\end{equation}
We insert this expression in Eq.(\ref{phitau}) and take care of the
constants 
\begin{equation}
\phi \left( x,y\right) =-2\phi _{0}\sum_{n=1}^{N}\sum_{l=-\infty }^{\infty }%
\frac{1}{\left[ \gamma \left( y-y_{n}\right) -ilD\right] ^{2}}  \label{phiin}
\end{equation}

\section{The dynamics of the singularities}

\subsection{Extending by inclusion of the $x$ coordinate}

The $x$ dependence in this equation comes from the dependence on the
variable $x$ of the position $y_{n}\left( x\right) $ of the poles in the
complex $y$ plane. Now we impose that this form of $\phi $ verifies the
equation. 
\begin{equation*}
\frac{\partial ^{2}\phi }{\partial x^{2}}+\frac{\partial ^{2}\phi }{\partial
y^{2}}=\alpha \phi -\beta \phi ^{2}
\end{equation*}
\begin{equation*}
\frac{\partial \phi }{\partial x}=-4\gamma \phi
_{0}\sum_{n=1}^{N}\sum_{l=-\infty }^{\infty }\left( \frac{dy_{n}\left(
x\right) }{dx}\right) \frac{1}{\left[ \gamma \left( y-y_{n}\right) -ilD%
\right] ^{3}}
\end{equation*}
\begin{eqnarray*}
\frac{\partial ^{2}\phi }{\partial x^{2}} &=&-4\gamma \phi
_{0}\sum_{n=1}^{N}\sum_{l=-\infty }^{\infty }\left\{ \frac{1}{\left[ \gamma
\left( y-y_{n}\right) -ilD\right] ^{3}}\right. + \\
&&\left. +3\gamma \left( \frac{dy_{n}\left( x\right) }{dx}\right) ^{2}\frac{1%
}{\left[ \gamma \left( y-y_{n}\right) -ilD\right] ^{4}}\right\}
\end{eqnarray*}
\begin{equation*}
\frac{\partial \phi }{\partial y}=4\gamma \phi
_{0}\sum_{n=1}^{N}\sum_{l=-\infty }^{\infty }\frac{1}{\left[ \gamma \left(
y-y_{n}\right) -ilD\right] ^{3}}
\end{equation*}
\begin{equation*}
\frac{\partial ^{2}\phi }{\partial y^{2}}=-12\gamma ^{2}\phi
_{0}\sum_{n=1}^{N}\sum_{l=-\infty }^{\infty }\frac{1}{\left[ \gamma \left(
y-y_{n}\right) -ilD\right] ^{4}}
\end{equation*}

We replace these formulas and those for $\phi $ and $\phi ^{2}$ in the
equation. We will examine the neighborhood of one of the poles, taking 
\begin{equation*}
y=y_{p}\left( x\right) +\varepsilon
\end{equation*}
where $\varepsilon $ is a small quantity and $p$ is one of the $N$ poles .
Expanding all terms in the equation in $\varepsilon $ we equal to zero the
coefficients of the same powers of $\varepsilon $. This will give us the
equations we have to impose to $y_{n}\left( x\right) $.

We have, with $\varepsilon \rightarrow 0$, 
\begin{eqnarray*}
\phi &=&\frac{\left( -2\phi _{0}\right) }{\gamma ^{2}}\frac{1}{\varepsilon
^{2}}+ \\
&&+\left( -2\phi _{0}\right) \sum_{\substack{ l=-\infty  \\ l\neq 0}}%
^{\infty }\frac{1}{\left( -ilD\right) ^{2}}+\left( -2\phi _{0}\right) \sum 
_{\substack{ n=1  \\ n\neq p}}^{N}\sum_{l=-\infty }^{\infty }\frac{1}{\left[
\gamma \left( y_{p}-y_{n}\right) -ilD\right] ^{2}}
\end{eqnarray*}
\begin{eqnarray*}
\frac{\partial ^{2}\phi }{\partial x^{2}} &=&\left( -4\phi _{0}\gamma
\right) \left( \frac{d^{2}y_{p}\left( x\right) }{dx^{2}}\right) \frac{1}{%
\gamma ^{3}\varepsilon ^{3}}+ \\
&&+\left( -4\phi _{0}\gamma \right) \left( \frac{d^{2}y_{p}\left( x\right) }{%
dx^{2}}\right) \sum_{\substack{ l=-\infty  \\ l\neq 0}}^{\infty }\frac{1}{%
\left( -ilD\right) ^{3}}+ \\
&&+\left( -4\phi _{0}\gamma \right) \sum_{\substack{ n=1  \\ n\neq p}}%
^{N}\left( \frac{d^{2}y_{n}\left( x\right) }{dx^{2}}\right) \sum_{l=-\infty
}^{\infty }\frac{1}{\left[ \gamma \left( y_{p}-y_{n}\right) -ilD\right] ^{3}}%
+ \\
&&+\left( -12\phi _{0}\gamma ^{2}\right) \left( \frac{dy_{p}\left( x\right) 
}{dx}\right) ^{2}\frac{1}{\gamma ^{4}\varepsilon ^{4}}+ \\
&&+\left( -12\phi _{0}\gamma ^{2}\right) \left( \frac{dy_{p}\left( x\right) 
}{dx}\right) ^{2}\sum_{\substack{ l=-\infty  \\ l\neq 0}}^{\infty }\frac{1}{%
\left( -ilD\right) ^{4}}+ \\
&&+\left( -12\phi _{0}\gamma ^{2}\right) \sum_{\substack{ n=1  \\ n\neq p}}%
^{N}\left( \frac{dy_{n}\left( x\right) }{dx}\right) ^{2}\sum_{l=-\infty
}^{\infty }\frac{1}{\left[ \gamma \left( y_{p}-y_{n}\right) -ilD\right] ^{4}}
\end{eqnarray*}
\begin{eqnarray*}
\frac{\partial ^{2}\phi }{\partial y^{2}} &=&\left( -12\phi _{0}\gamma
^{2}\right) \frac{1}{\gamma ^{4}\varepsilon ^{4}}+ \\
&&+\left( -12\phi _{0}\gamma ^{2}\right) \sum_{\substack{ l=-\infty  \\ %
l\neq 0}}^{\infty }\frac{1}{\left( -ilD\right) ^{4}}+ \\
&&+\left( -12\phi _{0}\gamma ^{2}\right) \sum_{\substack{ n=1  \\ n\neq p}}%
^{N}\sum_{l=-\infty }^{\infty }\frac{1}{\left[ \gamma \left(
y_{p}-y_{n}\right) -ilD\right] ^{4}}
\end{eqnarray*}
\begin{eqnarray*}
\phi ^{2} &=&4\phi _{0}^{2}\left\{ \frac{1}{\gamma ^{4}\varepsilon ^{4}}%
\right. + \\
&&+\frac{1}{\gamma ^{2}\varepsilon ^{2}}\left[ 2\sum_{\substack{ l=-\infty 
\\ l\neq 0}}^{\infty }\frac{1}{\left( -ilD\right) ^{2}}+2\sum_{\substack{ %
n=1  \\ n\neq p}}^{N}\sum_{l=-\infty }^{\infty }\frac{1}{\left[ \gamma
\left( y_{p}-y_{n}\right) -ilD\right] ^{2}}\right] + \\
&&\left. +\left[ \sum_{\substack{ l=-\infty  \\ l\neq 0}}^{\infty }\frac{1}{%
\left( -ilD\right) ^{2}}+\sum_{\substack{ n=1  \\ n\neq p}}%
^{N}\sum_{l=-\infty }^{\infty }\frac{1}{\left[ \gamma \left(
y_{p}-y_{n}\right) -ilD\right] ^{2}}\right] ^{2}\right\}
\end{eqnarray*}
We have to collect the terms containing the same powers of $\varepsilon $,
begining with the highest. For easy identification, we show separately the
contributions of each term of the original equation.

For $1/\varepsilon ^{4}$: 
\begin{eqnarray*}
&&\left( -12\phi _{0}\gamma ^{2}\right) \left( \frac{dy_{p}\left( x\right) }{%
dx}\right) ^{2}\frac{1}{\gamma ^{4}\varepsilon ^{4}}+ \\
&&\left( -12\phi _{0}\gamma ^{2}\right) \frac{1}{\gamma ^{4}\varepsilon ^{4}}%
+ \\
&&\left( -\alpha \right) \times 0+ \\
&&\beta 4\phi _{0}^{2}\frac{1}{\gamma ^{4}\varepsilon ^{4}} \\
&=&0
\end{eqnarray*}
This gives the equation 
\begin{equation}
\left( \frac{dy_{p}\left( x\right) }{dx}\right) ^{2}=\frac{1}{3}\frac{\phi
_{0}\beta }{\gamma ^{2}}-1  \label{protoec1}
\end{equation}
or 
\begin{equation}
\left( \frac{dy_{p}\left( x\right) }{dx}\right) ^{2}=1  \label{ec1}
\end{equation}

For $1/\varepsilon ^{3}$: 
\begin{eqnarray*}
&&\left( -4\phi _{0}\gamma \right) \left( \frac{d^{2}y_{p}\left( x\right) }{%
dx^{2}}\right) \frac{1}{\gamma ^{3}\varepsilon ^{3}}+ \\
&&0+ \\
&&\left( -\alpha \right) \times 0+ \\
&&\beta \times 0 \\
&=&0
\end{eqnarray*}
The resulting equation is 
\begin{equation}
\frac{d^{2}y_{p}\left( x\right) }{dx^{2}}=0  \label{ec2}
\end{equation}
Finally, the coefficient of $1/\varepsilon ^{2}$: 
\begin{eqnarray*}
&&0+ \\
&&0+ \\
&&\left( -\alpha \right) \times \frac{\left( -2\phi _{0}\right) }{\gamma ^{2}%
}\frac{1}{\varepsilon ^{2}}+ \\
&&\beta \times 4\phi _{0}^{2}\frac{1}{\gamma ^{2}\varepsilon ^{2}}\left[
2\sum_{\substack{ l=-\infty  \\ l\neq 0}}^{\infty }\frac{1}{\left(
-ilD\right) ^{2}}+2\sum_{\substack{ n=1  \\ n\neq p}}^{N}\sum_{l=-\infty
}^{\infty }\frac{1}{\left[ \gamma \left( y_{p}-y_{n}\right) -ilD\right] ^{2}}%
\right] \\
&=&0
\end{eqnarray*}
This results in the following constraint 
\begin{eqnarray}
\sum_{\substack{ l=-\infty  \\ l\neq 0}}^{\infty }\frac{1}{\left(
-ilD\right) ^{2}}+\sum_{\substack{ n=1  \\ n\neq p}}^{N}\sum_{l=-\infty
}^{\infty }\frac{1}{\left[ \gamma \left( y_{p}-y_{n}\right) -ilD\right] ^{2}}
&=&-\frac{\alpha }{2\phi _{0}\beta }  \label{cons} \\
&=&-\frac{1}{3}  \notag
\end{eqnarray}

From the Eqs.(\ref{ec2}) and (\ref{ec1}) it results that the
``trajectories'' \ $y_{p}\left( x\right) $ are linear on $x$. We have 
\begin{equation}
\frac{dy_{p}\left( x\right) }{dx}=\pm 1  \label{yp}
\end{equation}
or 
\begin{equation}
y_{p}\left( x\right) =\pm x+c_{p}  \label{ypsol}
\end{equation}
where $c_{p}$ are constants.

\subsection{Discussion of the constraint equation}

We have obtained the equations and the constraint using an expansion around
a singularity $y_{p}$. The choice is arbitrary and we can repeat the
calculations using another singularity. Obviously, the form of the equations
will not be changed. Finally, we will obtain a number of constraints equal
to the number of singularities.

A simpler form of the constraint can be written. 
\begin{equation*}
\sum_{\substack{ l=-\infty  \\ l\neq 0}}^{\infty }\frac{1}{\left(
-ilD\right) ^{2}}=\frac{1}{\left( -iD\right) ^{2}}2\sum_{l=1}^{\infty }\frac{%
1}{l^{2}}=-\frac{\pi ^{2}}{3D^{2}}
\end{equation*}
We introduce the notation 
\begin{equation*}
z_{pn}\equiv \frac{i\gamma }{D}\left( y_{p}-y_{n}\right)
\end{equation*}
and we have 
\begin{equation*}
-\frac{1}{D^{2}}\sum_{\substack{ n=1  \\ n\neq p}}^{N}\sum_{l=-\infty
}^{\infty }\frac{1}{\left( z_{pn}+l\right) ^{2}}=\frac{\pi ^{2}}{3D^{2}}-%
\frac{1}{3}
\end{equation*}
The second sum can be written 
\begin{equation*}
\sum_{l=-\infty }^{\infty }\frac{1}{\left( z_{pn}+l\right) ^{2}}=-\frac{1}{%
z_{pn}^{2}}+\psi ^{\prime }\left( z_{pn}\right) +\psi ^{\prime }\left(
-z_{pn}\right)
\end{equation*}
and the constraints become, for $p=1,...,N$%
\begin{equation*}
\sum_{\substack{ n=1  \\ n\neq p}}^{N}\left[ \frac{D^{2}}{\gamma ^{2}\left(
y_{p}-y_{n}\right) ^{2}}+\psi ^{\prime }\left( \frac{y_{p}-y_{n}}{iD/\gamma }%
\right) +\psi ^{\prime }\left( \frac{y_{n}-y_{p}}{iD/\gamma }\right) \right]
=-\frac{\pi ^{2}-D^{2}}{3}
\end{equation*}
In these formulas, $\psi ^{\prime }$ is the first derivative of the Euler 
\emph{psi}-function. This form can be useful if one wants to examine
numerically the validity of a particular choice of constants $c_{p}$.

\bigskip

Alternatively we can use the expansion for the square of the $\mathrm{cosech}
$ function, as before. 
\begin{equation*}
\sum_{l=-\infty }^{\infty }\frac{1}{\left[ \gamma \left( y_{p}-y_{n}\right)
-ilD\right] ^{2}}=\frac{\pi ^{2}}{D^{2}}\mathrm{cosech}^{2}\left[ \frac{\pi
\gamma }{D}\left( y_{p}-y_{n}\right) \right]
\end{equation*}
and the form of the constraint becomes 
\begin{equation}
\frac{\pi ^{2}}{D^{2}}\sum_{\substack{ n=1  \\ n\neq p}}^{N}\mathrm{cosech}%
^{2}\left[ \frac{\pi \gamma }{D}\left( y_{p}-y_{n}\right) \right] =\frac{\pi
^{2}}{3D^{2}}-\frac{1}{3}  \label{const2}
\end{equation}
The following identity exists 
\begin{equation}
\sum_{k=1}^{s-1}\mathrm{cosec}^{2}\left( \frac{k\pi }{s}\right) =\frac{%
s^{2}-1}{3}  \label{ident1}
\end{equation}
Comparing our equation with the identity and using the elementary relation 
\begin{equation*}
\mathrm{cosech}^{2}\left( ix\right) =-\mathrm{cosec}^{2}\left( x\right)
\end{equation*}
it is suggested to identify 
\begin{equation}
s\equiv \frac{D}{\pi }  \label{sdpi}
\end{equation}
\begin{equation}
\gamma \left( y_{p}-y_{n}\right) /i=k\pi  \label{kpi}
\end{equation}

At this moment the solution can be written 
\begin{eqnarray*}
\phi \left( x,y\right) &=&-2\phi _{0}\sum_{n=1}^{N}\sum_{l=-\infty }^{\infty
}\frac{1}{\left[ \gamma \left( y-y_{n}\right) -ilD\right] ^{2}} \\
&=&-2\phi _{0}\frac{\pi ^{2}}{D^{2}}\sum_{n=1}^{N}\mathrm{cosech}^{2}\left[ 
\frac{\pi \gamma }{D}\left( y\pm x+c_{n}\right) \right]
\end{eqnarray*}

\bigskip

Regarding the determination of the constants $c_{p}$, $p=1,...,N$ we note
that the condition (\ref{kpi}) only says that the differences between the
values must be a multiple of $\pi i$. However, there is the additional
constraint that the constants cannot vanish. This is because the function
would present singularities in the real space variable. On the other hand,
the singularities must be placed symmetrically along the imaginary axis.
These makes three restriction to any choice of the constants of integration

\begin{enumerate}
\item  the initial constraint, Eq.(\ref{const2}) which has been trasformed
into Eq.(\ref{kpi})

\item  the restriction $c_{p}\neq 0$

\item  the symmetrical positions along the imaginary axis, in order to have
real solutions.
\end{enumerate}

This leads to the following choice 
\begin{equation*}
\gamma c_{k}=k\pi i+\frac{i\pi }{2}
\end{equation*}
where $k=-N,...,N$.

We will assume in the following that the number of poles is infinite, 
\begin{equation*}
N\rightarrow \infty
\end{equation*}
However this will be discussed in more detail below.

The problem of ennumeration of poles with $\pm x$ is so suppressed but we
will have to make all steps separately for the two\ $\pm $ families. The
fact that we can choose only two families is related to the impossibility to
satisfy the constraints in the case when we would choose arbitrary
combinations of the sign of $x$'s in $\left( y_{p}-y_{n}\right) $.

The constraints become two infinite systems of equations implying only $%
c_{k} $, the constants. Then the solution can be written 
\begin{eqnarray*}
\phi \left( x,y\right) &=&-2\phi _{0}\frac{\pi ^{2}}{D^{2}}%
\sum_{n=1}^{\infty }\mathrm{cosech}^{2}\left[ \frac{\pi \gamma }{D}\left(
y+x+c_{n}\right) \right] + \\
&&-2\phi _{0}\frac{\pi ^{2}}{D^{2}}\sum_{n=1}^{\infty }\mathrm{cosech}^{2}%
\left[ \frac{\pi \gamma }{D}\left( y-x+c_{n}^{\prime }\right) \right]
\end{eqnarray*}
We can add to this expression the constraint equation, multiplied with $%
\left( 2\phi _{0}\right) $ and written as an identity with zero. 
\begin{eqnarray*}
\phi \left( x,y\right) &=& \\
&&\hspace*{-1.5cm}-2\phi _{0}\frac{\pi ^{2}}{D^{2}}\sum_{n=1}^{\infty }%
\mathrm{cosech}^{2}\left[ \frac{\pi \gamma }{D}\left( y+x+c_{n}\right) %
\right] + \\
&&\hspace*{-1.5cm}+\left( 2\phi _{0}\right) \frac{\pi ^{2}}{D^{2}}%
\sum_{n=1}^{\infty }\mathrm{cosech}^{2}\left[ \frac{\pi \gamma }{D}\left(
c_{p}-c_{n}\right) \right] -\left( 2\phi _{0}\right) \left( \frac{\pi ^{2}}{%
3D^{2}}-\frac{1}{3}\right) \\
&&\hspace*{-1.5cm}-2\phi _{0}\frac{\pi ^{2}}{D^{2}}\sum_{n=1}^{\infty }%
\mathrm{cosech}^{2}\left[ \frac{\pi \gamma }{D}\left( y-x+c_{n}^{\prime
}\right) \right] \\
&&\hspace*{-1.5cm}+\left( 2\phi _{0}\right) \frac{\pi ^{2}}{D^{2}}%
\sum_{n=1}^{\infty }\mathrm{cosech}^{2}\left[ \frac{\pi \gamma }{D}\left(
c_{p}^{\prime }-c_{n}^{\prime }\right) \right] -\left( 2\phi _{0}\right)
\left( \frac{\pi ^{2}}{3D^{2}}-\frac{1}{3}\right)
\end{eqnarray*}
If the constants are chosen as suggested before then we have to exclude from
the sum the term corresponding to $n=0$. 
\begin{eqnarray*}
\phi \left( x,y\right) &=& \\
&&\hspace*{-1.5cm}-2\phi _{0}\frac{\pi ^{2}}{D^{2}}\sum_{\substack{ %
n=-\infty  \\ n\neq 0}}^{\infty }\left\{ \mathrm{cosech}^{2}\left[ \frac{\pi
\gamma }{D}\left( y+x+\frac{i\pi }{2\gamma }+\frac{n\pi i}{\gamma }\right) %
\right] -\mathrm{cosech}^{2}\left[ \frac{\pi }{D}n\pi i\right] \right\} \\
&&\hspace*{-1.5cm}-\left( 2\phi _{0}\right) \left( \frac{\pi ^{2}}{3D^{2}}-%
\frac{1}{3}\right) -2\phi _{0}\frac{\pi ^{2}}{D^{2}}\mathrm{cosech}^{2}\left[
\frac{\pi \gamma }{D}\left( y+x+\frac{i\pi }{2\gamma }\right) \right] \\
&&\hspace*{-1.5cm}-2\phi _{0}\frac{\pi ^{2}}{D^{2}}\sum_{\substack{ %
n=-\infty  \\ n\neq 0}}^{\infty }\left\{ \mathrm{cosech}^{2}\left[ \frac{\pi
\gamma }{D}\left( y-x+\frac{i\pi }{2\gamma }+\frac{n\pi i}{\gamma }\right) %
\right] -\mathrm{cosech}^{2}\left[ \frac{\pi }{D}n\pi i\right] \right\} \\
&&\hspace*{-1.5cm}-\left( 2\phi _{0}\right) \left( \frac{\pi ^{2}}{3D^{2}}-%
\frac{1}{3}\right) -2\phi _{0}\frac{\pi ^{2}}{D^{2}}\mathrm{cosech}^{2}\left[
\frac{\pi \gamma }{D}\left( y-x+\frac{i\pi }{2\gamma }\right) \right]
\end{eqnarray*}
For a reason that will become clear later we can only chose the positive
sign of $i\pi /2$.

This can further be written 
\begin{eqnarray*}
\phi \left( x,y\right) &=& \\
&&\hspace*{-1.5cm}-2\phi _{0}\frac{\pi ^{2}}{D^{2}}\left( \frac{1}{3}+\sum 
_{\substack{ n=-\infty  \\ n\neq 0}}^{\infty }\left\{ \mathrm{cosech}^{2}%
\left[ \frac{\pi \gamma }{D}\left( y+x+\frac{i\pi }{2\gamma }+\frac{n\pi i}{%
\gamma }\right) \right] -\mathrm{cosech}^{2}\left[ \frac{\pi }{D}n\pi i%
\right] \right\} \right) \\
&&\hspace*{-1.5cm}-2\phi _{0}\frac{\pi ^{2}}{D^{2}}\mathrm{cosech}^{2}\left[ 
\frac{\pi \gamma }{D}\left( y+x+\frac{i\pi }{2\gamma }\right) \right] \\
&&\hspace*{-1.5cm}-2\phi _{0}\frac{\pi ^{2}}{D^{2}}\left( \frac{1}{3}+\sum 
_{\substack{ n=-\infty  \\ n\neq 0}}^{\infty }\left\{ \mathrm{cosech}^{2}%
\left[ \frac{\pi \gamma }{D}\left( y-x+\frac{i\pi }{2\gamma }+\frac{n\pi i}{%
\gamma }\right) \right] -\mathrm{cosech}^{2}\left[ \frac{\pi }{D}n\pi i%
\right] \right\} \right) \\
&&\hspace*{-1.5cm}-2\phi _{0}\frac{\pi ^{2}}{D^{2}}\mathrm{cosech}^{2}\left[ 
\frac{\pi \gamma }{D}\left( y-x+\frac{i\pi }{2\gamma }\right) \right] \\
&&\hspace*{-1.5cm}+\frac{4}{3}\phi _{0}
\end{eqnarray*}

Now we have to recall the identity for the doubly periodic elliptic
Weierstrass function $\wp $ 
\begin{eqnarray}
&&\frac{1}{3}+\mathrm{cosech}^{2}\left( \frac{\pi w}{L_{2}}\right) +
\label{ident2} \\
&&+\sum_{\substack{ n=-\infty  \\ n\neq 0}}^{\infty }\left\{ \mathrm{cosech}%
^{2}\left[ \frac{\pi }{L_{2}}\left( w+nL_{1}\right) \right] -\mathrm{cosech}%
^{2}\left[ \frac{n\pi L_{1}}{L_{2}}\right] \right\}  \notag \\
&=&\wp \left( w\right) \left( \frac{L_{2}}{\pi }\right) ^{2}  \notag
\end{eqnarray}
Where $L_{1}$ and $L_{2}$ are respectively the period on the real axis and
the period on the imaginary axis of the argument of $\wp $: $2\omega
_{1}=L_{1}$ and $2\omega _{2}=iL_{2}$. We can identify 
\begin{eqnarray*}
L_{1} &=&\pi i \\
L_{2} &=&D
\end{eqnarray*}
and the two variables 
\begin{eqnarray}
w &=&\gamma \left( y+x\right) +\frac{i\pi }{2}  \label{exw} \\
\text{or}\;w &=&\gamma \left( y-x\right) +\frac{i\pi }{2}  \notag
\end{eqnarray}

At this moment an important comment should be done. As we have seen the
solution of the constraint equation can be obtained on the basis of the
comparison with the identity Eq.(\ref{ident1}). We observe that the upper
limit of the summation in (\ref{ident1}) is interpreted as the number of
poles, 
\begin{equation*}
s-1=N
\end{equation*}
On the other hand we were led to identify 
\begin{equation*}
s=\frac{D}{\pi }
\end{equation*}
which means that the number of poles is the integer part 
\begin{equation*}
N=\left[ \frac{D}{\pi }\right] +1
\end{equation*}
Since we consider an infinite number of poles, $N\rightarrow \infty $, this
means that $D$ is infinite. This is necessary since we later use the
identity (\ref{ident2}) where the summation is extended over an infinite
number of integer values $n$. Since however we find that $L_{2}=D$, this
implies that the period $L_{2}$ of the Weierstrass function $\wp $ is
infinite. Certainly this is not acceptable in this particular approach, (but
in general it is meaningful) and this shows that the identification using
Eq.(\ref{ident2}) can only be an approximation. This also means that the
solution can only be approximative. However, if we decide to use this
approximative identification, we have to understand in what consists this
approximation and where we can expect to intervene the error we have
introduced by that.

We first note that the number of terms which is required in (\ref{ident2})
to obtain a good approximation of the Weierstrass function is not necessarly
large, for a significant area in the perodicity parallelogram of the complex
argument. This means that actually $D$ can be taken finite and in this case
the use of Eq.(\ref{ident1}) becomes legitimate. What is the number of
poles, $N$ and accordingly the length of periodicity $L_{1}=D$ in a
reasonable approximation? From numerical experience it may be accepted that
about five terms in the summation determining the Weierstrass function gives
a reasonable result. Then the number of poles retained in the sums (\emph{%
i.e.} $s-1=N$) can be a few units and this also means that $D\sim $few units.

In the following we will write the solution as being expressed in terms of
the Weierstrass function, but we have to remember that it is the result of
an approximation. It can be written 
\begin{eqnarray*}
\phi \left( x,y\right) &=&\frac{4}{3}\phi _{0} \\
&&-2\phi _{0}\frac{\pi ^{2}}{D^{2}}\wp \left[ \gamma \left( x+y\right) +%
\frac{i\pi }{2}\right] \left( \frac{D}{\pi }\right) ^{2} \\
&&-2\phi _{0}\frac{\pi ^{2}}{D^{2}}\wp \left[ \gamma \left( x-y\right) +%
\frac{i\pi }{2}\right] \left( \frac{D}{\pi }\right) ^{2}
\end{eqnarray*}
or 
\begin{equation}
\phi \left( x,y\right) =\frac{4}{3}\phi _{0}-2\phi _{0}\left\{ \wp \left[
\gamma \left( x+y\right) +\frac{i\pi }{2}\right] +\wp \left[ \gamma \left(
x-y\right) +\frac{i\pi }{2}\right] \right\}  \label{weipm}
\end{equation}
This expression represents in plane a system of rectangular cells.

We will prove later that a single family of poles corresponding to one of
the two possibilities $\pm x$ in Eq.(\ref{ypsol}) , which generates either
the first or the second term in the Eq.(\ref{weipm}) can provide an \emph{%
exact} solution to the equation. This justifies the calculations presented
in this section, since after them we are led directly to the form of the
solution.

\section{The r\^{o}le of the elliptic function}

\subsection{Basic information on the Weierstrass elliptic function
(lemniscate type)}

The most important consequence of the calculations that have been presented
is the generation of a \emph{periodic} solution and the confirmation, by the
method of motion of poles, of the class of quasi-one-dimensional solutions.
In the absence of a systematic integration procedure (like Inverse
Scattering Transform) we can be at least sure that this class represents a
significant part of the space of solutions.

Even if Eq.(\ref{weipm}) is the result of an approximation, it clearly shows
that the solution should be searched in a form of a double periodic elliptic
Weierstrass function. This function has appeared in our calculations
independently of any attempt to integrate a one-dimensional version of the
equation.

We note that 
\begin{eqnarray*}
\frac{\partial ^{2}}{\partial x^{2}}\wp \left[ \gamma \left( x+y\right) +%
\frac{i\pi }{2}\right] &=&6\gamma ^{2}\left( \wp \left[ \gamma \left(
x+y\right) +\frac{i\pi }{2}\right] \right) ^{2}-\frac{g_{2}}{2} \\
&=&\frac{\partial ^{2}}{\partial y^{2}}\wp \left[ \gamma \left( x+y\right) +%
\frac{i\pi }{2}\right]
\end{eqnarray*}
Here $g_{2}$ is the second coefficient in the standard expression for the
definition of the Weierstrass function 
\begin{equation*}
\wp ^{-1}\left( y\right) \equiv u=\int_{y}^{\infty }\frac{dt}{\sqrt{%
4t^{3}-g_{2}t-g_{3}}}
\end{equation*}

\bigskip

In the following we will use this information to derive an \emph{exact}
solution of the Petviashvilli equation 
\begin{equation}
\Delta \phi =\alpha \phi -\beta \phi ^{2}  \label{petv2}
\end{equation}

We start by a linear substitution 
\begin{equation}
\phi \left( x,y\right) =s\psi \left( x,y\right) +\frac{\alpha }{2\beta }
\label{linsubst}
\end{equation}
(where $s$ is a constant) which transform the equation into 
\begin{equation}
\Delta \psi =-s\beta \psi ^{2}+\frac{\alpha ^{2}}{4s\beta }-\frac{1}{s}%
\Delta \left( \frac{\alpha }{2\beta }\right)  \label{delpsi}
\end{equation}
We have allowed formally the space variation of the coefficients. For an
easier reference to the properties of the Weierstrass function, we make a
change of variables 
\begin{eqnarray}
x &\rightarrow &x_{c}=ix  \label{xyim} \\
y &\rightarrow &y_{c}=iy  \notag
\end{eqnarray}
which gives 
\begin{equation}
\Delta _{c}\psi \left( x_{c},y_{c}\right) =s\beta \psi \left(
x_{c},y_{c}\right) -\frac{\alpha ^{2}}{4s\beta }+\frac{1}{s}\Delta \left( 
\frac{\alpha }{2\beta }\right)  \label{psic}
\end{equation}

We look for a solution having the dependence on the coordinates $x$ and $y$
mediated by a new function, which we denote by $u\left( x_{c},y_{c}\right) $%
\begin{equation}
\psi \equiv \psi \left( u\right)  \label{psideu}
\end{equation}
with 
\begin{equation}
u\equiv u\left( x_{c},y_{c}\right)  \label{udexcyc}
\end{equation}
and now we proceed to express the equation (\ref{psic}) using this form. 
\begin{eqnarray*}
\frac{\partial \psi }{\partial x_{c}} &=&\frac{d\psi }{du}\frac{\partial u}{%
\partial x_{c}} \\
\frac{\partial ^{2}\psi }{\partial x_{c}^{2}} &=&\frac{d^{2}\psi }{du^{2}}%
\left( \frac{\partial u}{\partial x_{c}}\right) ^{2}+\frac{d\psi }{du}\frac{%
\partial ^{2}u}{\partial x_{c}^{2}}
\end{eqnarray*}
and analogous for $y$. 
\begin{eqnarray*}
\Delta _{c}\psi \left( x_{c},y_{c}\right) &=&\frac{d^{2}\psi }{du^{2}}\left[
\left( \frac{\partial u}{\partial x_{c}}\right) ^{2}+\left( \frac{\partial u%
}{\partial y_{c}}\right) ^{2}\right] \\
&&+\frac{d\psi }{du}\left[ \Delta _{c}u\right] \\
&=&s\beta \psi ^{2}-\frac{\alpha ^{2}}{4s\beta }+\frac{1}{s}\Delta \left( 
\frac{\alpha }{2\beta }\right)
\end{eqnarray*}
Suppose we find a function $u\left( x_{c},y_{c}\right) $ verifying the
conditions 
\begin{eqnarray}
\Delta _{c}u &=&0  \label{condu} \\
\left( \frac{\partial u}{\partial x_{c}}\right) ^{2}+\left( \frac{\partial u%
}{\partial y_{c}}\right) ^{2} &=&q  \notag
\end{eqnarray}
where $q$ is a constant. In this case the equation could be written 
\begin{equation}
\frac{d^{2}\psi }{du^{2}}=\frac{s\beta }{q}\psi ^{2}-\frac{\alpha ^{2}}{%
4s\beta q}+\frac{1}{qs}\Delta \left( \frac{\alpha }{2\beta }\right)
\label{ecpsi}
\end{equation}
and this form has a known solution, from the identifications 
\begin{eqnarray}
\psi \left( u\right) &\equiv &\wp \left( u\right)  \label{idwei} \\
\frac{s\beta }{q} &=&6  \notag \\
\frac{\alpha ^{2}}{4s\beta q}-\frac{1}{qs}\Delta \left( \frac{\alpha }{%
2\beta }\right) &=&\frac{g_{2}}{2}  \notag
\end{eqnarray}
This is because we have the known relationship for the Weierstrass function 
\begin{equation}
\frac{d^{2}\wp \left( u\right) }{du^{2}}=6\wp ^{2}\left( u\right) -\frac{%
g_{2}}{2}  \label{wppp}
\end{equation}
which has exactly the same form as (\ref{ecpsi}). We find from Eq.(\ref
{idwei}) 
\begin{equation}
q=\frac{s\beta }{6}  \label{forq}
\end{equation}
\begin{equation}
g_{2}=\frac{3\alpha ^{2}}{\left( s\beta \right) ^{2}}-\frac{6}{s^{2}\beta }%
\Delta \left( \frac{\alpha }{\beta }\right)  \label{forg2}
\end{equation}

\section{Exact periodic solution of the Flierl-Petviashvilli equation}

We now turn to the solutions of the constraint equations defining $u$. It is
clear that we may chose 
\begin{equation}
u\left( x_{c},y_{c}\right) =ay_{c}+bx_{c}+\tau  \label{uform}
\end{equation}
where $a$, $b$ and $\tau $ can be complex. The following condition results 
\begin{equation*}
a^{2}+b^{2}=\frac{s\beta }{6}
\end{equation*}
In addition, a choice of $a$ and $b$ real numbers, which makes the first two
terms in (\ref{uform}) purely imaginary, should be coroborated with the
suggestion from Eq.(\ref{exw}) where the poles resulted shifted with a
symmetric quantity with respect to the real axis. This time, due to the
change of variables (\ref{xyim}) everything is rotated. This yields the
choice for $\tau $ as half of the period $\left( 2\omega \right) $ on the
real axis 
\begin{equation*}
\tau =\frac{\left( 2\omega \right) }{2}
\end{equation*}
and the final form of the solution to the equation (\ref{petv2}) 
\begin{equation*}
\phi \left( x,y\right) =\frac{\alpha }{2\beta }+s\wp \left( iay+ibx+\omega
|g_{2}=\frac{3\alpha ^{2}}{\left( s\beta \right) ^{2}}+\frac{6}{s^{2}\beta }%
\Delta \left( \frac{\alpha }{\beta }\right) ,g_{3}\right)
\end{equation*}
The second Weierstrass coefficient $g_{3}$ is left unspecified but it must
be \emph{constant}.

We conclude that an exact solution to the Petviashvilli equation with \emph{%
constant} coefficients $\alpha $ and $\beta $ is 
\begin{equation}
\phi \left( x,y\right) =\frac{\alpha }{2\beta }+s\wp \left( iay+ibx+\omega
|g_{2}=\frac{3\alpha ^{2}}{\left( s\beta \right) ^{2}}\right)  \label{sol}
\end{equation}
with the condition 
\begin{equation}
a^{2}+b^{2}=\frac{s\beta }{6}  \label{ab}
\end{equation}

Now we understand the nature of the choice which is implicitely done when we
discuss the solutions consisting of arbitrarly rotated ridges issued by
translating one-dimensional profiles. It means to remain in the class of
functions $u$ whose Laplacean is zero and the gradient is constant,
according to the conditions (\ref{condu}). Or, it can easily be seen that
the only functions that verify these conditions have the form of linear
combinations of the variables $x$ and $y$.

\section{The physical relevance of this periodic solution}

As a mathematical result, Eq.(\ref{sol}) is the \emph{exact} response to the
problem of solving the Eq.(\ref{eqorig}). This is valid for any sign of $%
\alpha $ and $\beta $ since the function $\wp $ is doubly periodic. The
physical problem, as usual, is more complicated. The coefficients $\alpha $
and $\beta $ which we have assumed constants actually have a certain space
variation. We think however that it is worth examining the consequences of
this solution for the nonlinear plasma models.

In order to discuss possible physical applications of this solution we have
to handle easily its numerical form. Although we have assumed that the two
parameters $\alpha $ and $\beta $ are \emph{constants}, we will make
estimations on the base of their physical origin. For this we remind that
the basic physical constants are (Lakhin et al. \cite{Lakhin}, Spatscheck 
\cite{Spatschek1}, Horton and Hasegawa \cite{HH}) 
\begin{eqnarray}
\alpha &=&\frac{1}{\rho _{s}^{2}}\left( 1-\frac{v_{\ast }}{u}\right)
\label{alphabeta} \\
\beta &=&\frac{T_{e}}{2u^{2}eB_{0}^{2}\rho _{s}^{2}}\frac{\partial }{%
\partial x}\left( \frac{1}{L_{n}}\right) =\frac{e}{2m_{i}u^{2}}\frac{%
\partial }{\partial x}\left( \frac{1}{L_{n}}\right)  \notag
\end{eqnarray}
The coefficient $\alpha $ has the dimension $\left( length\right) ^{-2}$ as
it should. We introduce the following normalization for the potential 
\begin{equation*}
\phi \rightarrow \frac{e\phi }{T_{e}}
\end{equation*}
which only affects the coefficient $\beta $%
\begin{eqnarray*}
\beta &\rightarrow &\frac{e}{2m_{i}u^{2}}\frac{\partial }{\partial x}\left( 
\frac{1}{L_{n}}\right) \frac{T_{e}}{e} \\
&=&\frac{T_{e}}{2m_{i}u^{2}}\frac{\partial }{\partial x}\left( \frac{1}{L_{n}%
}\right) \\
&=&\frac{c_{s}^{2}}{2u^{2}}\frac{\partial }{\partial x}\left( \frac{1}{L_{n}}%
\right)
\end{eqnarray*}
At this moment the units are

\begin{itemize}
\item  in the first term: $\Delta $ is measured in $\left( m^{-2}\right) $;
the potential is adimensionalised, it is $\phi \rightarrow e\phi /T_{e}$ for
which we have the order of magnitude 
\begin{equation*}
\frac{e\phi }{T_{e}}\sim \frac{\widetilde{n}}{n_{0}}
\end{equation*}

\item  the second term: the coefficient $\alpha $ is in $\left( m\right)
^{-2}$; the potential is adimensionalised.

\item  the third term: the coefficient $\beta $ is in $\left( m\right) ^{-2}$%
; the potential is adimensionalised.
\end{itemize}

With these values coefficients we have to calculate 
\begin{equation}
a^{2}+b^{2}=\frac{s\beta }{6}  \label{a2b2}
\end{equation}
\begin{equation}
g_{2}=\frac{3\alpha ^{2}}{\left( s\beta \right) ^{2}}+\frac{6}{s^{2}\beta }%
\Delta \left( \frac{\alpha }{\beta }\right)  \label{g2exp}
\end{equation}
The latter parameter is connected with the periodicity lengths $L_{1}$ and $%
iL_{2}$ or half-periods $\left( \omega ,\omega ^{\prime }\right) $ of the
Weierstrass function \cite{Byrd}. 
\begin{eqnarray}
\omega &=&\int_{e_{1}}^{\infty }\frac{dt}{\sqrt{4t^{3}-g_{2}t-g_{3}}}%
=\int_{e_{1}}^{\infty }\frac{dt}{\sqrt{4\left( t-e_{1}\right) \left(
t-e_{2}\right) \left( t-e_{3}\right) }}  \label{omr} \\
&=&\int_{e_{3}}^{e_{2}}\frac{dt}{\sqrt{4\left( t-e_{1}\right) \left(
t-e_{2}\right) \left( t-e_{3}\right) }}  \notag \\
&=&\frac{K}{\sqrt{e_{1}-e_{3}}}  \notag
\end{eqnarray}
The other period is 
\begin{eqnarray}
\omega ^{\prime } &=&\int_{e_{2}}^{e_{1}}\frac{dt}{\sqrt{4\left(
t-e_{1}\right) \left( t-e_{2}\right) \left( t-e_{3}\right) }}  \label{omi} \\
&=&\frac{K^{\prime }}{\sqrt{e_{1}-e_{3}}}  \notag
\end{eqnarray}
In these formulas 
\begin{equation}
k^{2}=\frac{e_{2}-e_{3}}{e_{1}-e_{3}}  \label{elk2}
\end{equation}
is the \emph{modulus of the Jacobian elliptic functions and integrals}. The
similar quantity 
\begin{equation}
k^{\prime }=\sqrt{1-k^{2}}  \label{elkp}
\end{equation}
is the \emph{complimentary modulus}. The quantity 
\begin{eqnarray}
K\left( k\right) &\equiv &K=F\left( \frac{\pi }{2},k\right)  \label{celintK}
\\
&=&\int_{0}^{\pi /2}\frac{d\theta }{\sqrt{1-k^{2}\sin ^{2}\theta }}  \notag
\end{eqnarray}
is the \emph{complete elliptic integral of the first kind}. And analogously 
\begin{equation}
K^{\prime }=K\left( k^{\prime }\right)  \label{celintKp}
\end{equation}

The order assumed is 
\begin{equation*}
e_{3}<e_{2}<e_{1}
\end{equation*}
As will become later more clear, our problem has led to a definition of the
Weierstrass function with only one coefficient , $g_{2}$ , fixed by the
conditions, while $g_{3}$ cannot be made precise. This is because the
equation we have used is the differential equation for the second
derivative, $\wp ^{\prime \prime }\left( u\right) $ and this equation only
depends on $g_{2}$.

The procedure for obtaining numerical results is :

\begin{itemize}
\item  assume physical parameters, density, temperature, etc. and calculate $%
\rho _{s}$, $v_{\ast }$, $u$, $\kappa _{n}\equiv L_{n}^{-1}$, $\partial
\kappa _{n}/\partial x$;

\item  calculate the coefficients of the original equation $\alpha $ and $%
\beta $ from Eqs.(\ref{alphabeta});

\item  calculate $g_{2}$ from Eq.(\ref{g2exp}) and chose the value of $g_{3}$%
. Find $e_{1},e_{2},e_{3}$;

\item  calculate the half-periods $\omega $ and $\omega ^{\prime }$ on the
real and imaginary axis from Eqs.(\ref{omr}) and (\ref{omi});

\item  chose values for $a$ and $b$ such as to verify the Eq.(\ref{a2b2});

\item  define the space region $\left( y,x\right) $ \emph{i.e.} (poloidal,
radial) with $y$ and $x$ measured in some typical Larmor radius $\rho _{s0}$;

\item  compute the complex variable: $iay+ibx$ and scale to $2\omega
^{\prime }$; take the real part of the argument as half the period on the
real axis, in order to have real solution.

\item  calculate the Weierstrass function of the complex argument, and
associate it with the point $\left( x,y\right) $, then calculate the
solution by multiplying with $s$ and adding the constant $\alpha /\left(
2\beta \right) $;
\end{itemize}

\subsection{An example}

We will assume aproximate values 
\begin{equation*}
\rho _{s}\sim 10^{-3}\;\left( m\right)
\end{equation*}
\begin{equation*}
1-\frac{v_{\ast }}{u}\sim 0.5
\end{equation*}
Then physical values for these quantities are 
\begin{equation*}
\alpha \sim \frac{1}{\left( 10^{-3}\right) ^{2}}\times 0.5=5\times
10^{5}\;\left( m^{-2}\right)
\end{equation*}
We take 
\begin{equation*}
\left( \frac{c_{s}}{u}\right) ^{2}\sim 10^{6}
\end{equation*}
and 
\begin{equation*}
\frac{\partial }{\partial x}\left( \frac{1}{L_{n}}\right) \sim 20\;\left(
m^{-2}\right)
\end{equation*}
then 
\begin{equation*}
\beta \sim 10^{7}\;\left( m^{-2}\right)
\end{equation*}

We see that we can multiply all terms with $\rho _{s0}^{2}\sim
10^{-6}\;\left( m^{-2}\right) $. In this way all distances will be expressed
in units of Larmor radius $\rho _{s0}$. It results 
\begin{eqnarray*}
\alpha &\rightarrow &\alpha =0.5 \\
\beta &\rightarrow &\beta =10
\end{eqnarray*}

The space variation of the \emph{physical} coefficients $\alpha $ and $\beta 
$ are only in the radial $x$ direction and is in general weak. The
contribution of this part to the estimated value of $g_{2}$ is approximately
one tenth from the first part (see also Appendix A)

We chose a factor of scale $s$ for the final amplitude of the part coming
from $\wp $ in the potential perturbation. This is a parameter that will
result in general from the physical initial conditions, together with $g_{3}$%
. We take 
\begin{equation*}
s=0.05
\end{equation*}

For the following calculations, we leave $g_{3}$ a free parameter. We use a
computer code that calculates the value of the Weierstrass function for any
complex argument, by reducing everything at a fundamental paralleogram with
sides equal with $1$ on the real as well as on the imaginary axis. This
means that the real part of the argument must be scaled with $\left( 2\omega
\right) $ and the imaginary part of the argument must be scaled with $\left(
2\omega ^{\prime }\right) $. After calculating the argument of the
Weierstrass function, the value to be inserted in the subroutine is 
\begin{equation*}
z=\frac{iay+ibx}{2\omega ^{\prime }}+\left( \frac{2\omega }{2}\right) \frac{1%
}{2\omega }
\end{equation*}
With these values the equations defining the parameters in our solution
becomes 
\begin{equation*}
a^{2}+b^{2}=0.09
\end{equation*}
\begin{equation*}
g_{2}=\frac{3\alpha ^{2}}{\left( s\beta \right) ^{2}}+\frac{6}{s^{2}\beta }%
\Delta \left( \frac{\alpha }{\beta }\right) \approx 3
\end{equation*}
We take 
\begin{equation*}
g_{3}=0
\end{equation*}
and find 
\begin{eqnarray}
e_{1} &=&0.866  \label{e123} \\
e_{2} &=&0  \notag \\
e_{3} &=&-0.866  \notag
\end{eqnarray}
\begin{eqnarray}
k &=&0.707  \label{kkp} \\
k^{\prime } &=&0.707  \notag
\end{eqnarray}
We find that the \emph{half-periods} of the Weierstrass function are 
\begin{eqnarray}
\omega &=&1.40879  \label{omomp} \\
\omega ^{\prime } &=&1.40879  \notag
\end{eqnarray}
We chose 
\begin{eqnarray}
a &=&0.0912  \label{abcho} \\
b &=&0.2738  \notag
\end{eqnarray}
and we have to scale on the real axis with 
\begin{equation}
2\omega =2.817  \label{doiom}
\end{equation}
and on the imaginary axis 
\begin{equation}
2\omega ^{\prime }=2.817  \label{doiomp}
\end{equation}
This gives a first estimation of the width of the layer along the $x$
direction: 
\begin{equation}
\delta x\sim \frac{2\omega ^{\prime }}{b}\sim 10  \label{delx}
\end{equation}
which means about $1\;cm$. We calculate the profile of the streamfunction $%
\phi \left( y,x\right) $ on a poloidal-radial domain of extension $\left(
40\rho _{s}\times 40\rho _{s}\right) $.

For example, we show the structure of the solution in the two plots. 
\begin{figure}[tbph]
\centerline{\includegraphics[height=10cm]{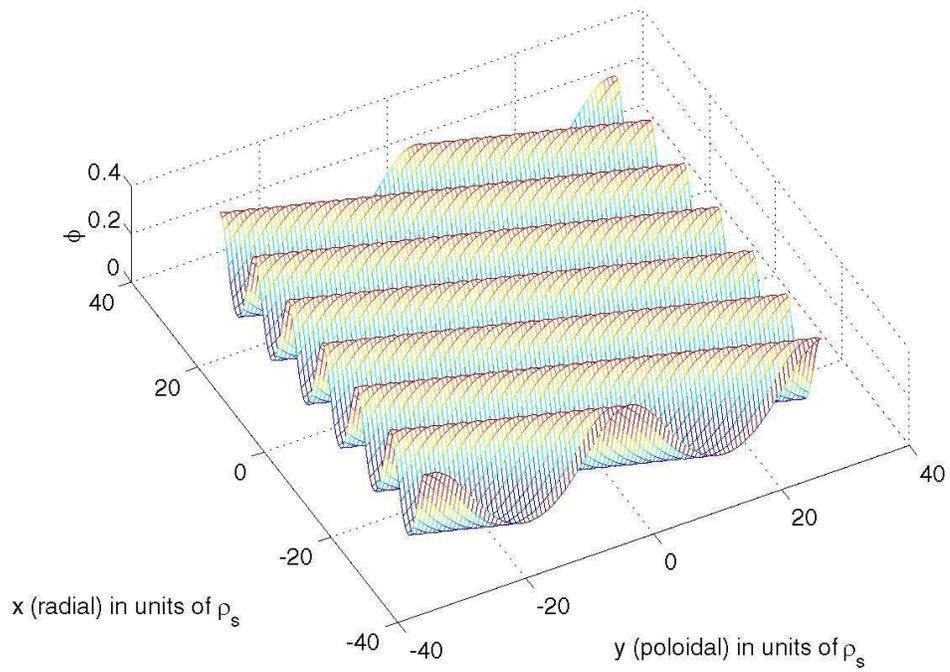}}
\caption{The streamfunction $\protect\phi $ solution of the Petviashvilli
equation for the ratio $\frac{b}{a}=0.9$.}
\label{fig1}
\end{figure}
\begin{figure}[tbph]
\centerline{\includegraphics[height=10cm]{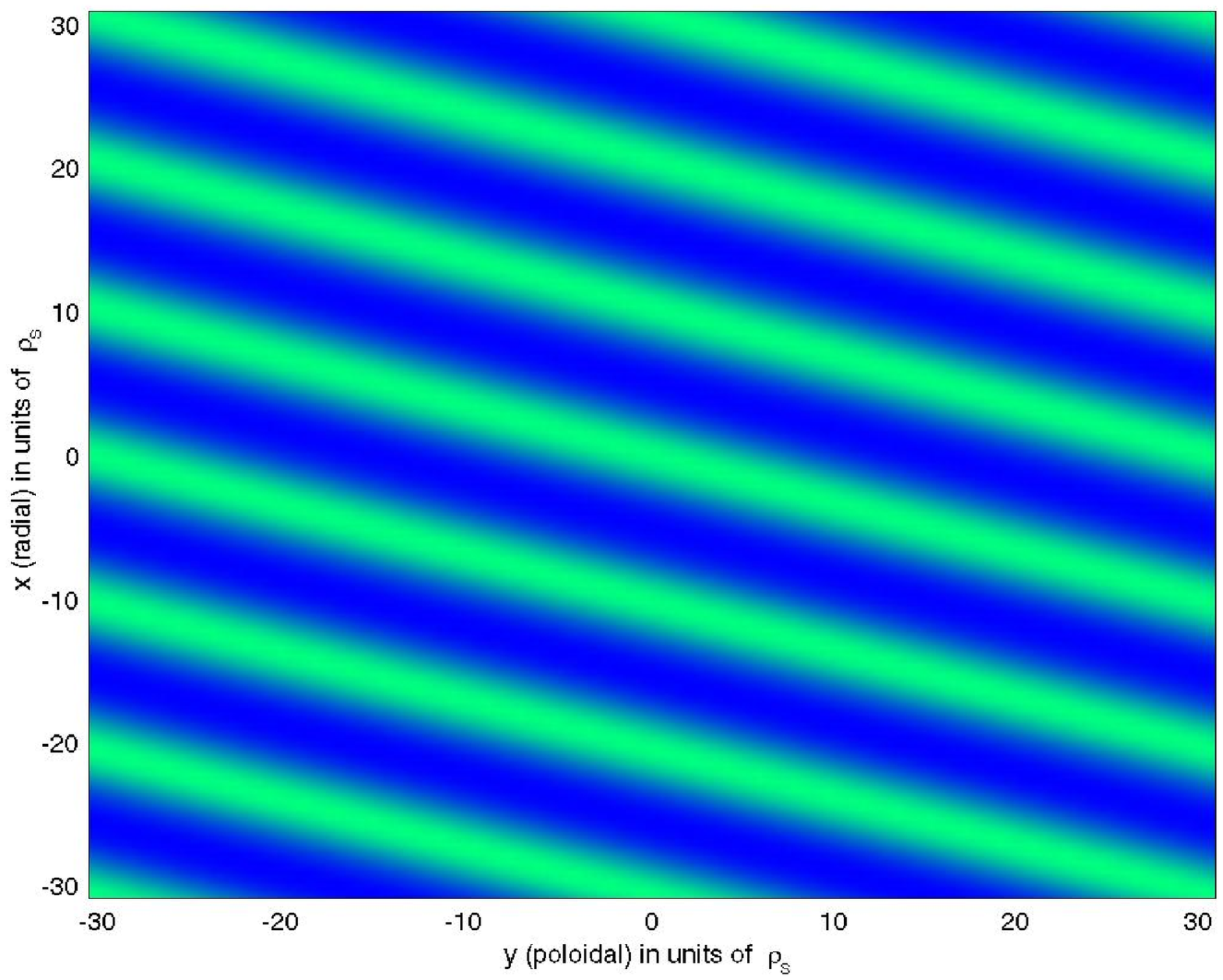}}
\caption{The same as Figure 1, in contour plot.}
\label{fig2}
\end{figure}

It is interesting to examine the structure of the flow field induced by the
potential as a function of the position along the minor radius in tokamak.
We first perform a numerical simulation using a one dimensional transport
code \cite{Florin1}. Details are given in Appendix B. The physical data from
simulation consists of 
\begin{equation*}
T_{e}\left( r\right) ,T_{i}\left( r\right) ,n\left( r\right) ,\rho
_{s}\left( r\right) ,c_{s}\left( r\right) ,L_{n}\left( r\right)
,L_{T_{e}}\left( r\right) ,v_{\ast }\left( r\right) 
\end{equation*}
which are transfered to a code that calculates the parameters $\alpha \left(
r\right) $ , $\beta \left( r\right) $, $a\left( r\right) $, $b\left(
r\right) $, $g_{2}\left( r\right) $ and, takes a fixed value for $g_{3}$.
Then the half-periods $\omega \left( r\right) $ and $\omega ^{\prime }\left(
r\right) $ are calculated. These parameters are represented in the figures
below. 
\begin{figure}[tbph]
\centerline{\includegraphics[height=19cm]{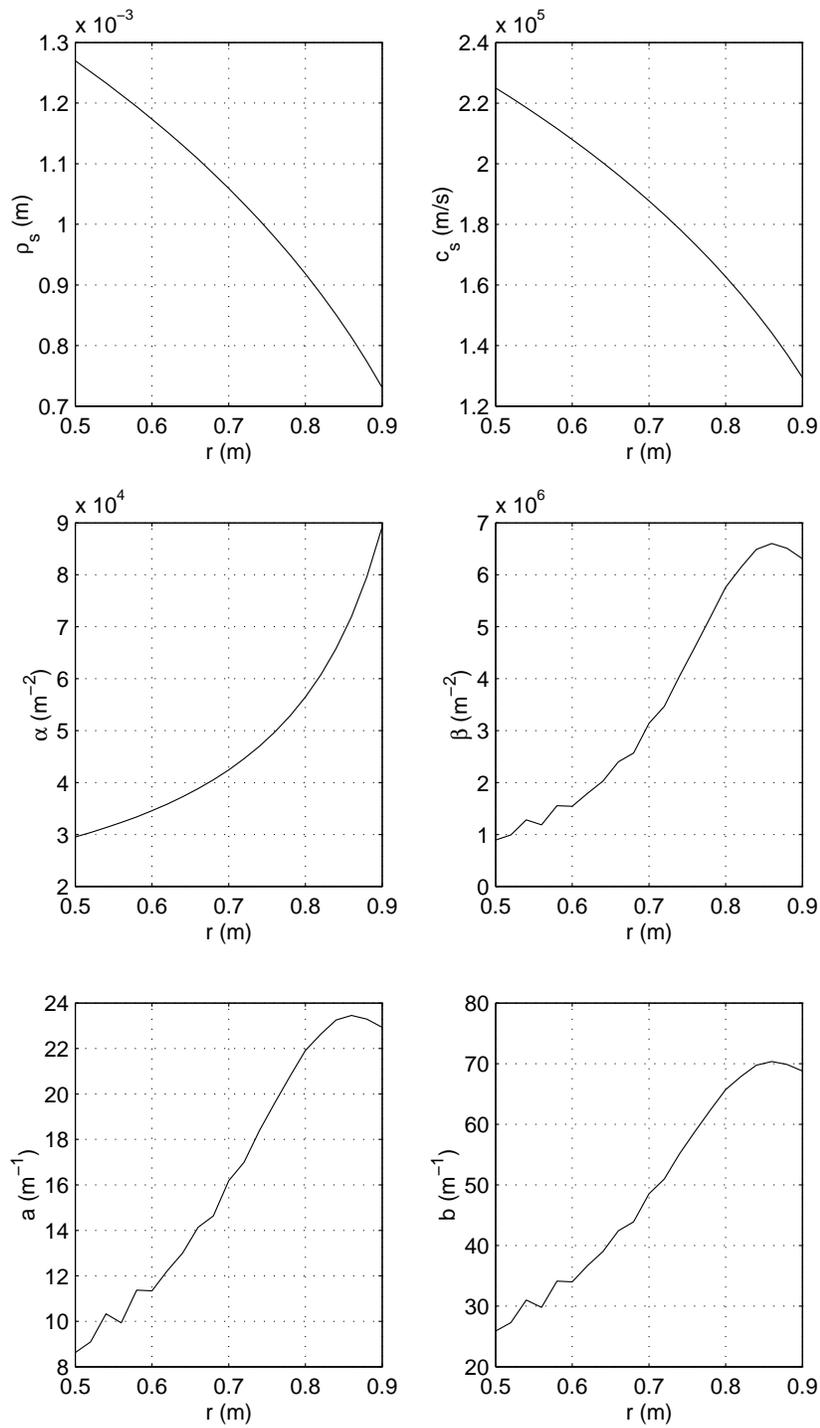}}
\caption{Physical parameters and the derived values $\protect\alpha $ , $%
\protect\beta $, $a$ and $b$ as function of $r$.}
\label{figparom}
\end{figure}
\begin{figure}[tbph]
\centerline{\includegraphics[height=19cm]{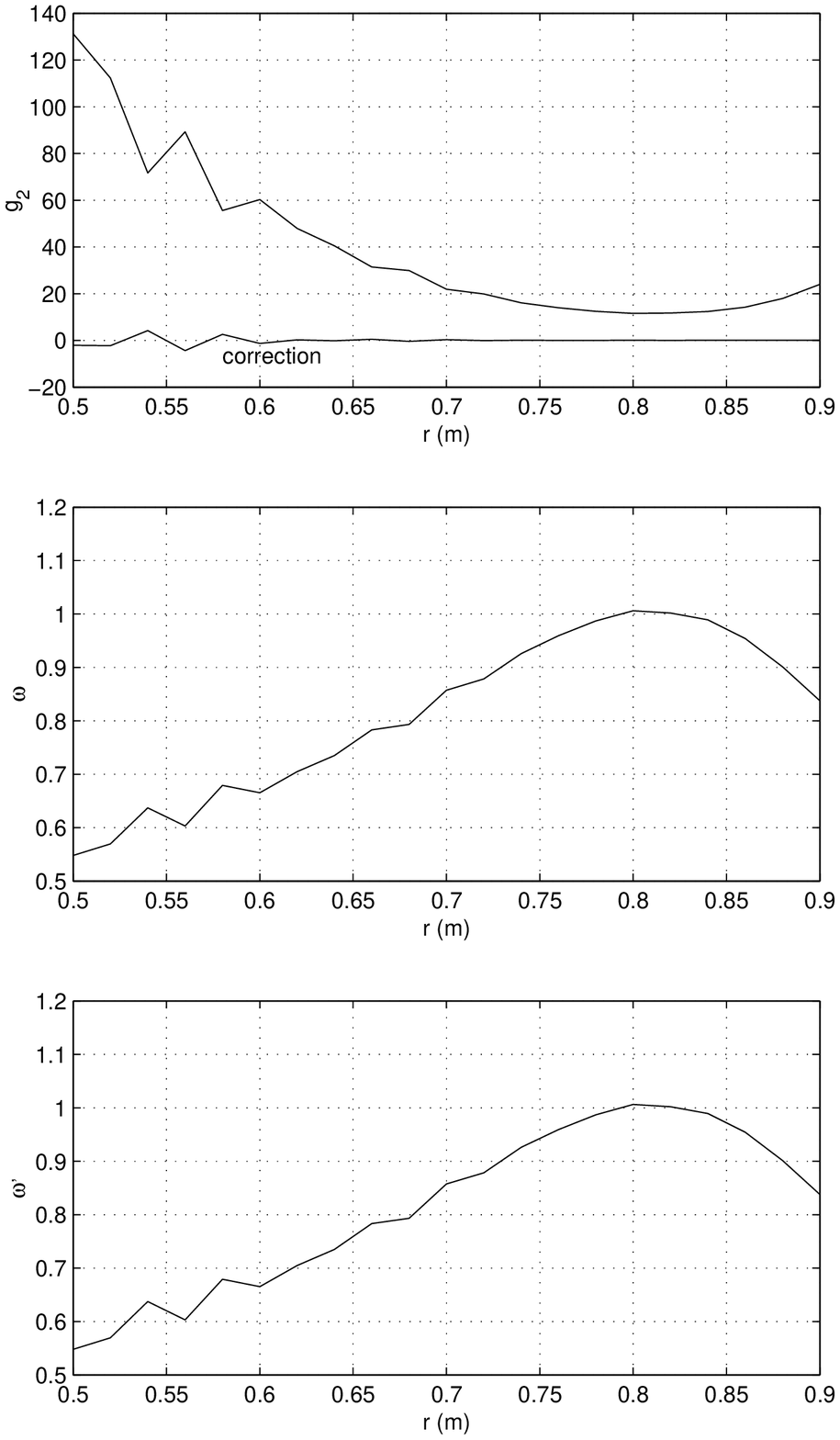}}
\caption{Plot of the parameters of the Weierstrass function. In the plot for 
$g_{2}$ the lower line represents the correction due to the space variation
of the physical parameters.}
\label{omep}
\end{figure}
\begin{figure}[tbph]
\centerline{\includegraphics[height=19cm]{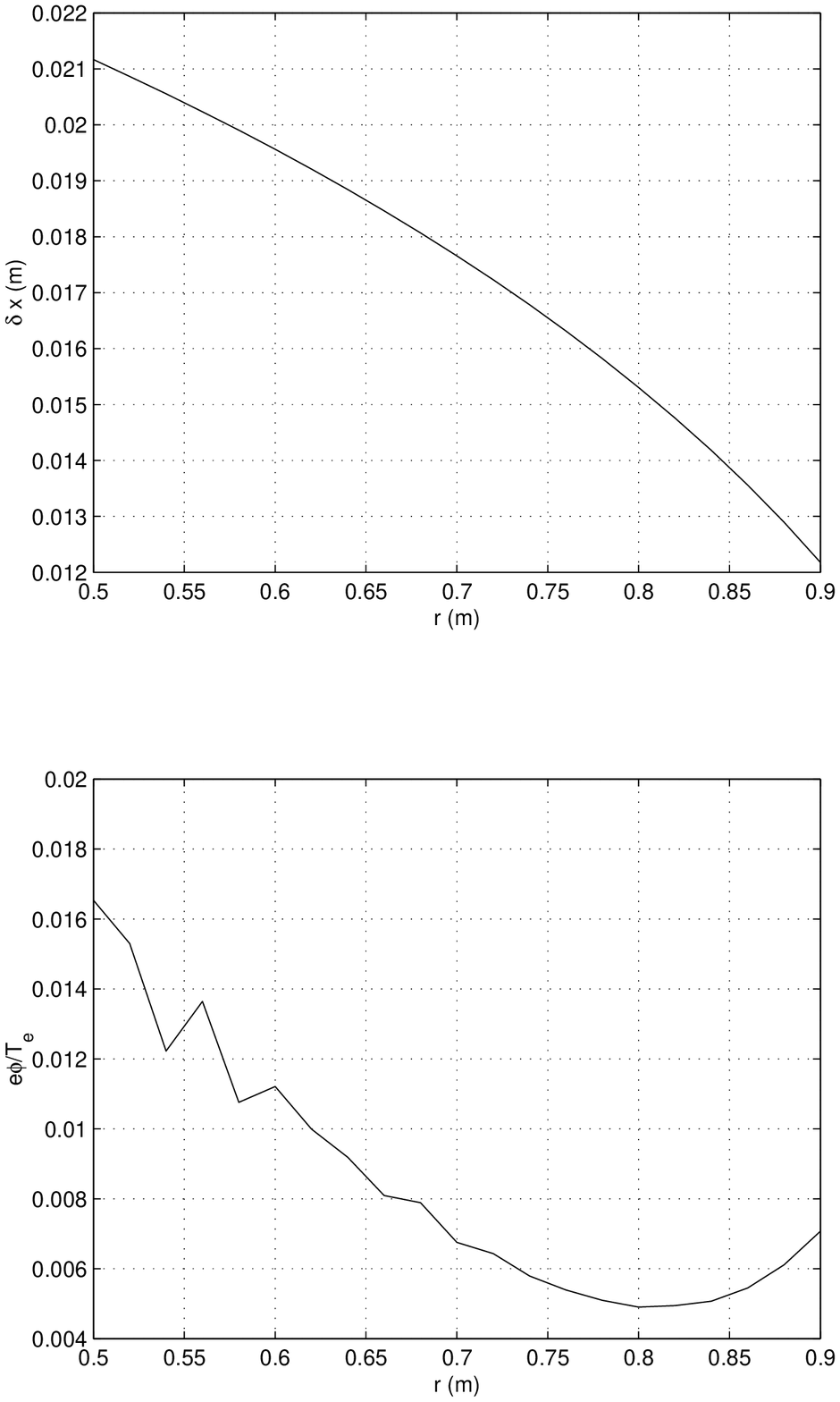}}
\caption{Width of the layer of flow and amplitude induced by the
perturbation.}
\label{figlayer}
\end{figure}
\begin{figure}[tbph]
\centerline{\includegraphics[height=8.5cm]{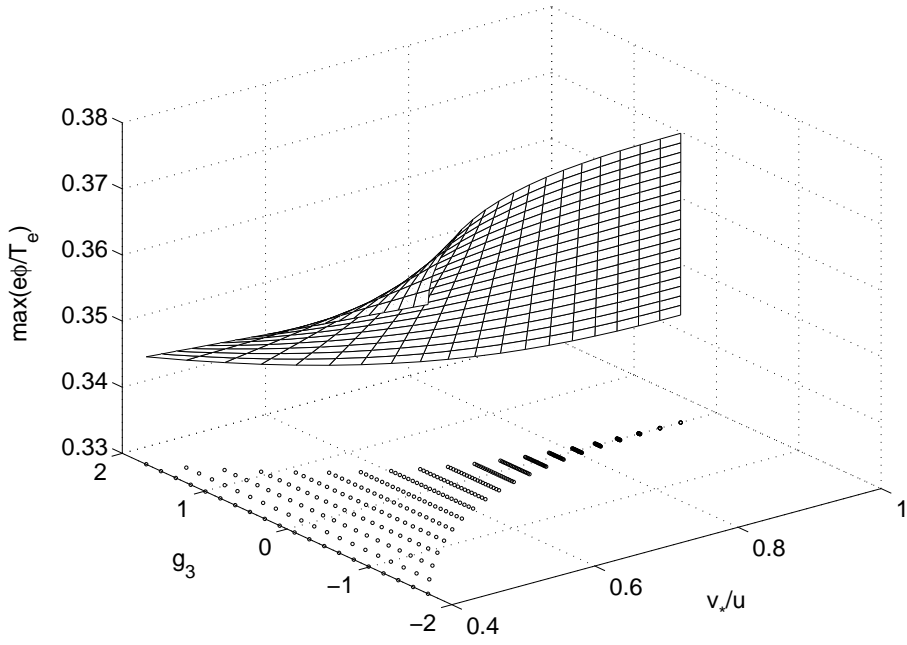}}
\caption{Variation of the amplitude $e\protect\phi /T_{e}$ with $g_{3}$ and $%
v_{\ast }/u$.}
\label{vamp}
\end{figure}
\begin{figure}[tbph]
\centerline{\includegraphics[height=8.5cm]{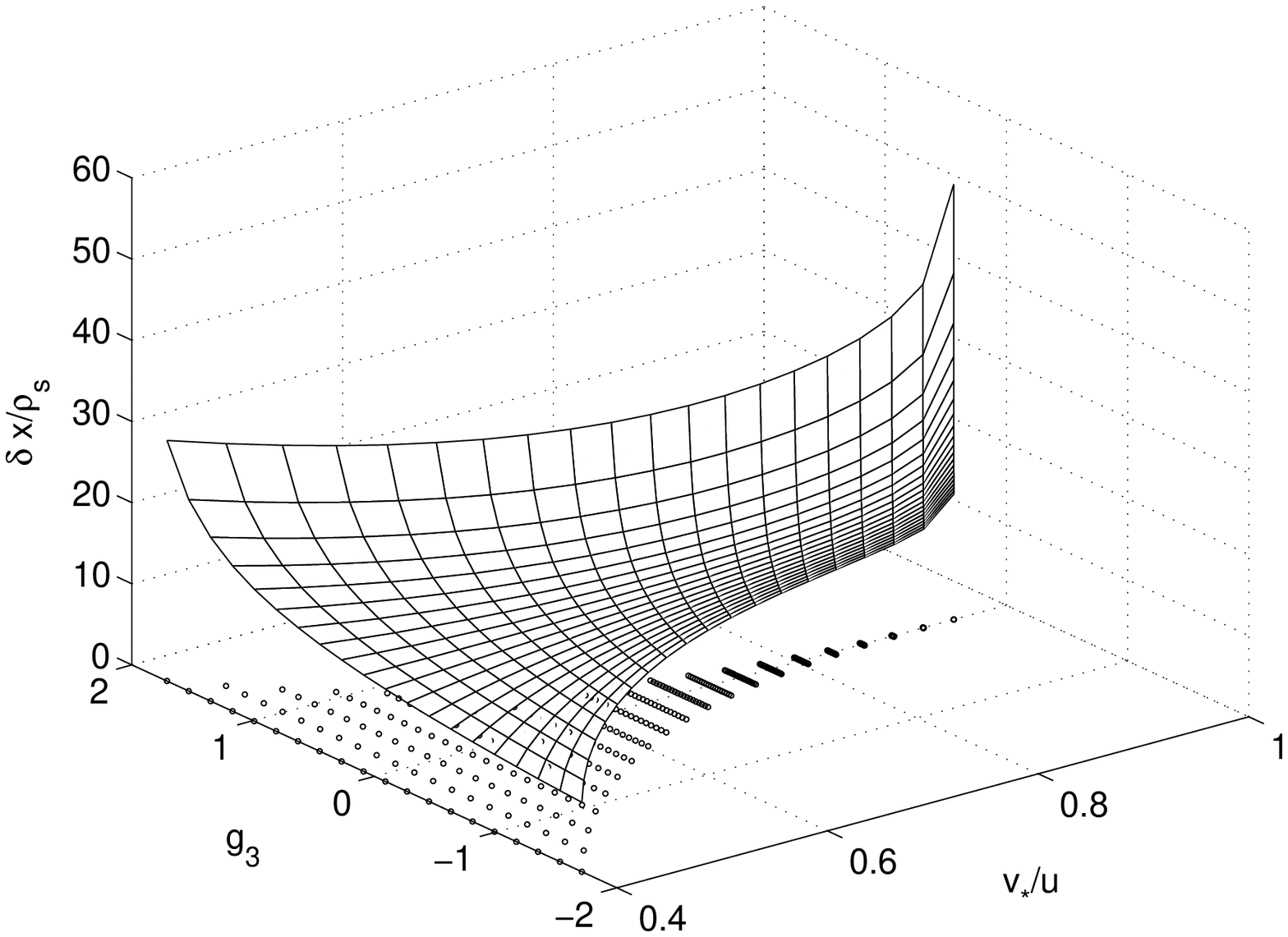}}
\caption{Variation of the width $\protect\delta x$ of a layer with $g_{3}$
and $v_{\ast }/u$.}
\label{vwid}
\end{figure}
We have studied the effect of the variation of the second Weierstrass
parameter, $g_{3}$. When it is taken equal to zero, we find that the width
of the layer of poloidal flow, Eq.(\ref{delx}) is proportional to the local
Larmor radius, 
\begin{equation*}
\delta x=3.924K\left( k^{\prime }\right) \left( 1-v_{\ast }/u\right)
^{-1/2}\rho _{s}
\end{equation*}
(with the constants of Eqs. (\ref{celintKp}) and (\ref{kkp})). The
variations of the width of a layer $\delta x/\rho _{s}$ and of the maximum
perturbation $e\phi /T_{e}$ with $g_{3}$ and $v_{\ast }/u$ are shown in Figs.%
\ref{vamp} and \ref{vwid}. On this graphs only the part where the
discriminant of the Weierstrass function is positive $\Delta \equiv
g_{2}^{3}-27g_{3}^{2}>0$ are shown, since there the roots $e_{1,2,3}$ are
all real (these points are also indicated in projection on the plane). The
characteristics of an initial physical field that can evolve toward this
layered geometry of flow can be inferred from these graphs by eliminating
the common variable $g_{3}$.

\section{Conclusion}

It has been known from the experiments on Rayleigh-Benard convection that at
higher Reynolds number there is a second bifurcation (the first being from
purely conducting to convective rolls phase) where the pattern of the fluid
flow exhibits tilted structures. Close to the upper and to the lower
boundaries there is a deformation of the rolls, the ``wind'' as has been
designated, which significantly enhances the Reynolds stress and favorizes
second instability and eventually global displacement of the fluid along the
boundaries. This finding, supported independently by suggestive results from
the numerical simulations of the ITG instability led to the idea that a
tilting instability may favorise, by an increase of the Reynolds stress, the
generation of zonal flow in plasma. The solutions we have found is a new
factor that should be included in this physical description.

Certainly, the tilt of the radially elongated eddies specific to the ITG
potential pattern can be favorable to the spontaneous generation of sheared
poloidal flow. On the other hand, the solution we have found has a strong
similarity to the tilted cells and zonal flows. It may be possible that
after a certain degree of tilting is attained, the plasma evolves
spontaneously to a solution of the type described above. This solution
intrinsically consists of layers of sheared flow. What is more important is
that this solution has all the usual attributes of an exactly integrable
structure: it is more robust and may represent an attractor. It can be
considered that relating the generation of the zonal flow to the process of
evolution of a system to a robust attractor is a useful approach to be
further examined.

In this work we have restricted to the nonlinear dynamics of the $2D$
structures. As a further extension of the nonlinear solution of
electrostatic drift waves described by the Hasegawa-Mima equation, the
electromagnetic $3D$ vortical motion has also been studied \cite{SW}, \cite
{WS}. In particular it has been shown that a $2D$ vortical motion in the $%
\left( r,\theta \right) $-plane propagates along the magnetic line of force.
Nonlinear coupling of drift-shear Alfven waves may lead to a novel nonlinear
solution. Recently this topic has been extended and the formulation has been
updated in Ref. \cite{MW}. These are however out of the scope of the present
work and will be a subject of future research.

\textbf{Acknowledgements}.

This work has been performed during the stay of two of the authors (M.O.V.
and F.S.) at NIFS as visiting professors. They acknowledge the hospitality
of Prof. M. Fujiwara and Prof. O. Motojima. This work is partially supported
by the Grant-in-Aid for Scientific Research of MEXT Japan and by the
collaboration programmes of the National Institute for Fusion Science and of
the Research Institute for Applied Mechanics of Kyushu University.


\section*{Appendix A : Discussion on the weak space variation of the
parameters}

\appendix
\renewcommand{\theequation}{A.\arabic{equation}} \setcounter{equation}{0}

In order to obtain the exact solution we had to assume that the coefficients
are constant. Later, we have considered a weak variation of the coefficients
with the coordinate $x$, induced by the presence of the physical parameters.
We can assume that this is a reasonable representation of the real situation
only if we have an adiabatic variation of $g_{2}$ with the physical
parameters.

The numerical study of the dependence of the elliptic function on the
physical parameters shows that $g_{2}$ is indeed slowly varying and that the
correction due to the second term in the expression of $g_{2}$ can be
neglected.

One can see that the weaker restriction can be formulated as follows: $g_{2}$
is independent on $u=iay+ibx+\omega $, which means that $g_{2}$ is constant
along the lines $ay+bx=const$ and can only have a slow parametric variation
in the direction transversal to this family of lines. Taking into account
the expression of $g_{2}$ and that $\alpha $ and $\beta $ has variation
mainly along the radial direction, we can reformulate the restriction by
requiring that $g_{2}$ has no variation along lines quasi-parallel to the
poloidal dierction, and can only have a weak dependence perpendicular on
these lines. This explains our choice of combination of $a$ and $b$ in the
numerical study presented above.

We will make a scaling transformation to reduce the Weierstrass function to
constant coefficient $g_{2}$. For this we recall the \emph{formula for
homogeneity} of the Weierstrass function 
\begin{equation}
\wp \left( u;g_{2},g_{3}\right) =\mu ^{2}\wp \left( \mu u;\frac{g_{2}}{\mu
^{4}},\frac{g_{3}}{\mu ^{6}}\right)  \label{homog}
\end{equation}
We take 
\begin{equation*}
g_{3}=0
\end{equation*}
\begin{eqnarray*}
\mu &=&g_{2}^{1/4} \\
&=&\left[ \frac{3\alpha ^{2}}{s^{2}\beta ^{2}}-\frac{6}{s^{2}\beta }\Delta
\left( \frac{\alpha }{\beta }\right) \right] ^{1/4}
\end{eqnarray*}
and the formula becomes 
\begin{equation}
\wp \left( u;g_{2},g_{3}\right) =g_{2}^{1/2}\wp \left(
g_{2}^{1/4}u;1,0\right)  \label{gampl}
\end{equation}
or 
\begin{equation}
\phi \left( x,y\right) =\frac{\alpha }{2\beta }+g_{2}^{1/2}s\wp \left(
g_{2}^{1/4}\left( iay+ibx+\omega \right) ;1,0\right)  \label{gphi}
\end{equation}

This formula may serve for a numerical investigation of the space variation
and the estimation of the error in using the solution based on constant
coefficients.

We ask the stronger condition, that $g_{2}$ is simply a constant 
\begin{equation*}
g_{2}=\frac{3\alpha ^{2}}{s^{2}\beta ^{2}}-\frac{6}{s^{2}\beta }\Delta
\left( \frac{\alpha }{\beta }\right) =\text{const}
\end{equation*}
This is essentially a differential equation which strongly constrain the
space dependence of the physical parameters. Taking for example const$=p$,
and remembering that the physical variation is on $x$, we have 
\begin{equation*}
\frac{d^{2}}{dx^{2}}\left( \frac{\alpha }{\beta }\right) =\frac{\alpha ^{2}}{%
2\beta }-\frac{\beta s^{2}p}{6}
\end{equation*}
Now we assume that the radial third derivative of the density is small and
take then $\beta $ constant. The equation for $\alpha $ is 
\begin{equation}
\alpha ^{\prime \prime }=6\alpha ^{2}-2\beta ^{2}s^{2}p  \label{eqalpha}
\end{equation}
after a dividing the variable $x$ with $\sqrt{12}$. This is again the
Weierstrass equation and the solution is $\wp \left( x\right) $ where $x$ is
measured from the surface where the periods are calculated. Looking for
non-periodic solutions we find 
\begin{equation}
\alpha \left( x\right) =1-\frac{v_{\ast }}{u}=\left( \frac{\beta ^{2}s^{2}p}{%
3}\right) ^{1/2}-\frac{3\left( \beta ^{2}s^{2}p/3\right) ^{1/2}}{\cosh ^{2}%
\left[ \sqrt{3}\left( \beta ^{2}s^{2}p/3\right) ^{1/4}\right] }
\label{alpsol}
\end{equation}
We see that the behaviour of the right hand side in Eq.(\ref{alpsol}) is
approximately linear in the variable $\left( \beta ^{2}s^{2}p/3\right)
^{1/2} $. Then we can expect a condition of the form 
\begin{equation*}
\frac{v_{\ast }}{u}\approx \beta s\left( \frac{g_{2}}{3}\right) ^{1/2}
\end{equation*}
or 
\begin{equation*}
\frac{\rho _{s}}{L_{n}}\approx \frac{c_{s}}{2u}\frac{d}{dx}\left( \frac{1}{%
L_{n}}\right) s\left( \frac{g_{2}}{3}\right) ^{1/2}
\end{equation*}
Here all distances, $x$ and $L_{n}$ have been normalized at a typical Larmor
radius, $\rho _{s0}$. This can be reduced to a condition on the density
variation with the minor radius 
\begin{equation*}
\frac{d}{dr}\left( \ln \frac{1}{L_{n}}\right) \approx \left[ s\left( \frac{%
g_{2}}{3}\right) ^{1/2}\frac{\Omega _{i}}{2u}\right] ^{-1}\sim 0.05
\end{equation*}
The conclusion is that the density gradient length has a variation of the
type 
\begin{equation*}
L_{n}\sim \exp \left( -0.05\frac{x}{\rho _{s0}}\right)
\end{equation*}
on an interval $0<x<100\rho _{s0}$ in the region where this solution exists.
The fast variation of the density is favorable to the validity of this
solution.


\section*{Appendix B : Numerical simulation for tokamak plasma parameter's
profiles}

\appendix
\renewcommand{\theequation}{B.\arabic{equation}} \setcounter{equation}{0}

The code \cite{Florin1} solves the balance equations for energy, density and
fields. The variables are: the electron and ion temperatures $T_{e}$, $T_{i}$%
; the current density $j$, the poloidal magnetic field $B_{\theta }$, the
toroidal electric field $E_{\varphi }$, the electron density $n_{e}$, the
radial pinch particle velocity $V_{r}$. The following equations are
discretized on a one-dimensional space mesh (on the small radius) and
evloved in time by a semi-implicit scheme.

\begin{eqnarray*}
\frac{3}{2}\frac{\partial }{\partial t}\left( n_{e}T_{e}\right) &=&-\frac{1}{%
r}\frac{\partial }{\partial r}\left( r\left( -n_{e}\chi _{e}\frac{\partial
T_{e}}{\partial r}+n_{e}VT_{e}\right) \right) +Ej \\
&&-3\frac{m_{e}}{m_{i}}n_{e}\frac{T_{e}-T_{i}}{\tau _{ei}}%
-P_{rad}-P_{ion}+P_{add}^{e}
\end{eqnarray*}

\begin{eqnarray*}
\frac{3}{2}\frac{\partial }{\partial t}\left( n_{i}T_{i}\right) &=&-\frac{1}{%
r}\frac{\partial }{\partial r}\left( r\left( -n_{i}\chi _{i}\frac{\partial
T_{i}}{\partial r}+n_{i}VT_{i}\right) \right) \\
&&+3\frac{m_{e}}{m_{i}}n_{e}\frac{T_{e}-T_{i}}{\tau _{ei}}-P_{cx}+P_{add}^{i}
\end{eqnarray*}

\begin{equation*}
j=\frac{1}{\mu _{0}}\frac{1}{r}\frac{\partial }{\partial r}\left( rB_{\theta
}\right)
\end{equation*}

\begin{equation*}
\frac{\partial B_{\theta }}{\partial t}=\frac{\partial E}{\partial r}
\end{equation*}

\begin{equation*}
E=\eta j
\end{equation*}

\begin{equation*}
\frac{\partial n_{e}}{\partial t}=\frac{1}{r}\frac{\partial }{\partial r}%
\left( rD\frac{\partial n_{e}}{\partial r}\right) +S_{ion}
\end{equation*}

\begin{equation*}
V=\frac{1}{n_{e}}\left( D\frac{\partial n_{e}}{\partial r}+V_{pinch}\right)
\end{equation*}
Neutral atoms as well as impurities (Carbon, Oxygen, Iron, Wolfram,
Molibden) are considered. The transport coefficients are those of the
Merejkhin-Mukhovatov model and the specific parameters of the tokamak
correspond to the Tore-Supra device. The run is extended over 20 seconds and
the stationarity is reached within few seconds. Then we extract radial
dependent plasma parameter from a time slice at about 13.5 seconds.

\end{document}